\documentclass[%
 aip,
 amsmath,amssymb,
 reprint,%
]{revtex4-1}

\usepackage{graphicx}% Include figure files
\usepackage{dcolumn}% Align table columns on decimal point
\usepackage{bm}% bold math
\usepackage[mathlines]{lineno}
\usepackage{amsthm}
\usepackage{amssymb}% Enable numbering of text and display math
\usepackage[table,xcdraw]{xcolor}
\usepackage[utf8]{inputenc}
\usepackage[T1]{fontenc}
\usepackage{mathptmx}
\usepackage{mathrsfs}
\usepackage{mathtools} 
\usepackage{etoolbox}
\usepackage{booktabs}
\usepackage[table,xcdraw]{xcolor}

\usepackage{siunitx}
\usepackage{booktabs}   % for toprule, bottomrule and midrule
\usepackage{multirow}   % in table
\usepackage[referable]{threeparttablex}

\makeatletter
\def\@email#1#2{%
 \endgroup
 \patchcmd{\titleblock@produce}
  {\frontmatter@RRAPformat}
  {\frontmatter@RRAPformat{\produce@RRAP{*#1\href{mailto:#2}{#2}}}\frontmatter@RRAPformat}
  {}{}
}%
\makeatother
\newcolumntype{x}{>{$}l<{$}}  % math-mode version of "l" column type
\newcolumntype{y}{>{$}c<{$}}  % math-mode version of "c" column type
\newcolumntype{z}{>{$}r<{$}}  % math-mode version of "r" column type

\newcommand{\txtd}{\text{d}}
\newcommand{\dr}{\text{d}\ensuremath{\mathbf{r}}}
\newcommand{\Ne}{\ensuremath{N_\text{e}}}

\newcommand{\aPE}{\ensuremath{|\text{PE}|}}
\newcommand{\occ}{\text{occ}}
\newcommand{\du}{\text{d}\ensuremath{u}}
\newcommand{\vext}{\ensuremath{v_\text{ext}}}
\newcommand{\nx}{n_\text{x}}
\newcommand{\dWcPE}{\ensuremath{\delta E^\text{PE}_\text{c}}}
\newcommand{\ccsd}{\text{CCSD}}
\newcommand{\lda}{\text{LDA}}

\newcommand{\exact}{\text{Exact}}
\newcommand{\nc}{n_\text{c}}
\newcommand{\nxc}{n_\text{xc}}
\newcommand{\Ex}{E_\text{x}}
\newcommand{\Ec}{E_\text{c}}
\newcommand{\Exc}{E_\text{xc}}
\newcommand{\xc}{\text{xc}}
\newcommand{\xcx}{\text{x}}
\newcommand{\xcc}{\text{c}}
\newcommand{\ext}{\text{ext}}
\newcommand{\boldr}{\ensuremath{\mathbf{r}}}
\newcommand{\boldR}{\ensuremath{\mathbf{R}}}
\newcommand{\boldu}{\ensuremath{\mathbf{u}}}

\newcommand{\Vee}
{\ensuremath{\hat{V}_\text{ee}}}

\newcommand{\ndep}{\ensuremath{[n]}}

\newcommand{\sysXCh}{\ensuremath{n^\lambda_\xc}}
\begin{document}
\preprint{AIP/123-QED}

\title{Capturing the electron-electron cusp with the coupling-constant averaged exchange--correlation hole: a case study for Hooke's atoms}
% Force line breaks with \\
\author{Lin Hou}
\affiliation{
Physics and Engineer Physics Department, Tulane University.
}
 
\author{Tom J. P. Irons}%
\affiliation{
School of Chemistry, University of Nottingham, University Park, Nottingham NG7 2RD, United Kingdom
}

\author{Yanyong Wang}
\affiliation{
Physics and Engineer Physics Department, Tulane University.
}

\author{James W. Furness}
\affiliation{
Physics and Engineer Physics Department, Tulane University.
}

\author{Andrew M. Teale}
\affiliation{
School of Chemistry, University of Nottingham, University Park, Nottingham NG7 2RD, United Kingdom
}
\affiliation{
Hylleraas Centre for Quantum Molecular Sciences, Department of Chemistry, University of Oslo, P.O. Box 1033, N-0315 Oslo, Norway
}

\author{Jianwei Sun}
\affiliation{
Physics and Engineer Physics Department, Tulane University.
}

%\date{\today}
\begin{abstract}
In density functional theory the exchange--correlation (XC) energy functional can be defined exactly through the coupling-constant ($\lambda$) averaged XC hole $\bar{n}_\xc(\boldr,\boldr')$, representing the probability depletion of finding an electron at $\boldr'$ due to an electron at $\boldr$. Accurate knowledge of $\bar{n}_\xc(\boldr,\boldr')$ has been crucial for developing various XC energy density functional approximations and understanding their performance for real molecules and materials. However, 
there are very few systems for which accurate XC holes have been calculated, since this requires evaluating the one- and two-particle reduced density matrices for a reference wave function over a range of $\lambda$ whilst the electron density remains fixed at the physical ($\lambda=1$) density. 
Although the coupled-cluster singles and doubles (CCSD) method can yield exact results for a two-electron system in the complete basis set limit, it cannot capture the electron-electron cusp with commonly used finite basis sets. In this study, focusing on the Hooke's atom as a two-electron model system for which certain analytic solutions are known, we examine the effect of this cusp error on the XC hole calculated using CCSD. 
The Lieb functional is calculated at a range of coupling constants to determine the $\lambda$-integrated XC hole. 
Our results indicate that, for the Hooke's atoms, the 
error introduced by the description of the electron-electron cusp using Gaussian basis sets at the CCSD level is negligible compared to the basis set incompleteness error.
The system-, angle- and coupling-constant-averaged XC hole is calculated using the same approach and provides a benchmark against which the Perdew-Burke-Ernzerhof (PBE) and local density approximation (LDA) XC hole models are assessed.

%$\langle\bar{n}_\xc(u)\rangle$ 
\end{abstract}



%\thanks{}



\maketitle %\maketitle must follow title, authors, abstract and \pacs

\section{Introduction}\label{sec:intro}

% The structure for the introduction should be:

% 1) DFT is widely used; exact in principle, approximated for XC. 

% 2) XC hole undergirds Exc; and early successful XC approximations were based on XC hole, e.g., PW91

% 3) XC hole is very important/fundamental for understanding performance of DFAs for different systems and properties.

% 4) Exact XC holes of systems are well defined, but very difficult to calculate. There are few such calculations, even for simple systems (He).

% 5) There are two challenges in calculating the exact or very accurate XC holes: a) FCI or high level electronic structure theories, which typically scale exponentially with electron numbers and computationally expensive, have to be used; b) the coupling contant that scales the electron-electron interaction and comes into the definition of XC hole.

% 6) Lieb maximization is a technique that could be used with high-level electronic structure theories (e.g., CCSD) to solve the above problems, but limited to small systems (e.g., H2, give the citation here). 

% 7) CCSD is exact for 2-electron systems in the basis set limit and therefore is used for H2 with the Lieb maximization. However, the basis set effect on the Exc and XC hole, in particular the cusp condition of C hole, is not known. Here, we try to study this with the Hooke's atoms. Briefly introduce the Hooke's atoms.

Due to its relatively low computational scaling combined with high accuracy in the study of electronic structure of many-body systems, density functional theory (DFT) has become the most widely used electronic structure method with an increasing range of applications in condensed-matter physics, quantum chemistry, and materials science. In principle, DFT is an exact method with which the ground-state energy and electron density can be computed, from which many important physical and chemical properties can be extracted~\cite{Kohn1964/PhysRev.136.B864}. In practice, approximations must be introduced to DFT to make it computationally useful; in the Kohn-Sham formulation of DFT (KS-DFT),~\cite{KohnSham1965} the exchange--correlation (XC) component of the energy which carries the many-electron effects must be approximated. Therefore it is the quality of the XC approximation that determines the quality of a DFT calculation in predicting the total energy and other ground-state properties of interest.

An exact expression for the XC energy can be obtained in terms of the electron density $n(\boldr)$ and the coupling-constant ($\lambda$) averaged XC hole density $\bar{n}_\xc(\boldr,\boldr')$ via their Coulomb interaction~\cite{ParrYang1989DT} as
\begin{align}
    E_{\text{xc}}\ndep = \frac{1}{2}\iint\text{d}\boldr \dr'\, \frac{ n(\boldr) \bar{n}_{\text{xc}}(\boldr,\boldr')}{ |\boldr-\boldr'| } \label{eq:Exc_xch_express}
\end{align}
where $\bar{n}_\xc(\boldr,\boldr')$ is the probability depletion of finding an electron at $\boldr'$, given an electron located at $\boldr$. $\bar{n}_\xc(\boldr,\boldr')$ is entirely attributed to quantum effects, which include the self-interaction correction, the Pauli exclusion principle (arising from the exchange symmetry of indistinguishable electrons), and the electron-electron correlation resulting from the Coulombic repulsion.~\cite{Perdew2003DFTprimer} The first two effects give rise to the exchange hole density $\nx(\boldr,\boldr')$, which is completely negative and independent of the coupling constant. The remaining quantum effects produce the correlation hole, which is defined by subtracting the exchange hole density from $\bar{n}_\xc(\boldr,\boldr')$ as $\bar{n}_\xcc(\boldr,\boldr')=\bar{n}_\xc(\boldr,\boldr') -n_\xcx(\boldr,\boldr')$, yielding the $\lambda$-averaged correlation hole.

Eq.~(\ref{eq:Exc_xch_express}) guarantees an accurate evaluation of XC energy if an accurate XC hole model is provided. Thus the quality of XC hole models underpin the XC energy approximation and play a fundamental role in understanding the quality and assessing the performance of a diverse range of density functional approximations (DFAs) when applied to different systems and properties. However, practical DFT calculations only require approximations of the XC energy, leading to a tendency to neglect the importance of XC holes in favor of directly modeling the XC energy.
This trend has led to there being relatively few XC holes studies. Notably, early successful DFAs such as the PW91 approximation of Perdew and Wang~\cite{PW91} were based on modeling the XC hole, and the construction of the strongly constrained and appropriately normed (SCAN) density functional was also grounded in the understanding of XC holes.~\cite{Sun2015/scan} Recently, there have been new DFA developments based on XC holes.~\cite{Burke2020/CPDFT}

Although being formally defined in Eq.~(\ref{eq:Exc_xch_express}), XC holes are challenging to evaluate accurately, contributing to the scarcity of the XC hole studies. There are two significant challenges associated with this: i) the XC hole has to be calculated for each coupling constant $\lambda$ to evaluate the coupling-constant integrated XC hole;
ii) high-level electronic structure methods are required to obtain accurate ground-state wave functions for each $\lambda$. These methods typically have high-rank polynomial scaling with system size and become computationally intractable for large systems. The Lieb optimization approach~\cite{WYang1983/Lie_opt} can address challenge i) by transforming the problem of finding the ground-state electron density of a $\lambda$-interacting system into maximizing the Lieb functional of the $\lambda$-dependent external potential,~\cite{Lieb1983} while keeping the electron density fixed. In combination with the coupled-cluster singles and doubles excitation method (CCSD), the Lieb optimization method has been applied to two-electron systems, such as the Helium isoelectronic series, with a focus on the XC energy.~\cite{Teale2009,Teale2010} The CCSD method is exact in the complete basis set limit, equivalent to the full configuration interaction (FCI) approach for two-electron systems.

However, the $\lambda$-averaged XC hole has not been studied using the Lieb optimization with a CCSD reference wave function, even for the simple two-electron systems. Therefore, it is currently not known how the basis set influences the quality of the calculated XC hole and the associated electron-electron cusp condition~\cite{kimball1973short,Davidson1976} of the correlation hole~\cite{burke1995real} when the coupling-constant averaged quantities are considered. 
The electron-electron cusp condition describes the behavior of a many-electron wave function when two anti-parallel electrons come infinitesimally close to each other, arising due to the singularity of the Coulomb repulsion at the coalescence point. This dynamical correlation effect at zero separation introduces non-smoothness into the many-body wave function, which cannot be effectively represented by orbital product expansion wave functions
~\cite{Alavi2018/cusp_inFCIQMC}. Increasing the basis set size can help reduce the cusp error, but this approach is limited by the unfavorable computational scaling of high-level electronic structure methods.

In this study, we examine the electron-electron cusp condition and basis set effects on the XC hole through the calculation of the Lieb functional at the CCSD level for a simple model system, namely the Hooke atom (Hookium). By introducing the harmonic-oscillator potential as the external potential in the Hamiltonian of a two-electron system, given in atomic units as
\begin{equation}
    \hat{H}=-\frac{1}{2} \nabla^2_1 +\frac{1}{2}k\boldr^2_1 -\frac{1}{2} \nabla^2_2 +\frac{1}{2}k \boldr^2_2+\frac{1}{\left|\boldr_1-\boldr_2\right|}, \label{eq:hookium_orgin}
\end{equation}
the resulting problem is one of the few examples of a two-electron system for which a series of exact solutions exist, in this case an infinite set of solutions corresponding to different harmonic confinement constants, $k$.~\cite{taut1993two,kais1993density} The Hookium atom is therefore a useful reference for evaluating XC hole models since the exact XC holes can be computed.  

We commence in Section~\ref{sec:theory_method} by providing an overview of the theoretical framework for computing the $\lambda$-dependent XC hole, the Lieb optimization method, and the electron-electron cusp condition in Coulombic systems, and the solvable Hookium model. Computational details are then discussed in Section~\ref{sec:computation}. In Section~\ref{sec:results_discuss} we examine the basis set effects and cusp condition effects on the XC hole calculated at the CCSD level at $\lambda=1$ (the physical system), for which the exact wave function solution is known. We then compare and benchmark the local density approximation (LDA) and Perdew-Burke-Ernzerhof (PBE) XC hole models with the coupling-constant averaged XC hole from Lieb optimizations at the CCSD level. System- and angle-averaged XC holes are calculated to enable direct comparison between the benchmark data and these simple desnsity-functional models. Finally, we conclude our work with a brief summary in Section~\ref{sec:conclusion}.

\section{Theory and methodology}\label{sec:theory_method}

\subsection{The exchange--correlation hole and the coupling constant}\label{subsec:XChole_theo}
In KS-DFT, the ground-state energy of a many-electron system in an external potential $v_\ext(\boldr)$ is obtained by mapping the interacting system of electrons to an auxiliary non-interacting system of electrons with the same electron density. The Schr{\"o}dinger equation for this auxiliary system can then be solved in a basis of one-electron orbitals.~\cite{KohnSham1965} The ground-state energy is thus expressed as a functional of the electron density $n(\boldr)$, which can be resolved into the sum of several contributions as
\begin{equation}\label{eq:KS-total}
    E\ndep = T_\text{s}\ndep + E_\text{H}\ndep + \Exc\ndep + \int\dr\, \vext(\boldr) n(\boldr)
\end{equation}
where $T_\text{s}$ is the non-interacting kinetic energy, which is evaluated exactly using the KS orbitals, and $E_\text{H}$ the classical electrostatic Hartree energy, which is evaluated exactly in terms of $n(\boldr)$. The only term in Eq.~(\ref{eq:KS-total}) which must be approximated is the XC energy $E_\xc\ndep$, which describes all of the many-electron effects in the system.

The KS non-interacting system may be linked to the physically-interacting system by continuously varying the strength of the electron-electron interaction between the non-interacting and physically-interacting limits by scaling the two-electron operator $\Vee$ by a coupling-constant $\lambda$ between zero and one. The electronic state evolves through a family of solutions to the $\lambda$-interacting Hamiltonian,
\begin{equation}\label{eq:H_lambda}
    \hat{H}_\lambda = \hat{T} + \lambda\Vee + \sum_i v_\lambda(\boldr_i),
\end{equation}
where $\hat{T}$ is the kinetic energy operator and $v_\lambda$ a modified external potential, thus estabilishing an adiabatic connection between the non-interacting and physically-interacting systems.~\cite{LangrethPerdew/adiabatic_connection} The modified external potential $v_\lambda$ is determined for each interaction strength such that the density remains constant at the physical ($\lambda=1$) density for all $\lambda$. Clearly, $v_\lambda$ reduces to the local KS potential $v_\text{s}$ when $\lambda=0$ and is equal to the physical external potential $\vext$ when $\lambda=1$. 

Supposing $\Psi_\lambda$ is the normalized ground-state many-electron wave function of the $\lambda$-interacting system with $\Ne$ electrons, the second-order reduced density matrix is expressed as~\cite{Davidson1976,McWeeny1960/advances_DMtheory}
\begin{align}
    n^\lambda_2\left(\boldr,\boldr'\right) &\equiv \Ne(\Ne-1)\sum_{\sigma_1,\cdots,\sigma_N} \int\dr_3 \cdots\int \dr_N \notag \\
    &\qquad\quad \left|\Psi_\lambda\left(\boldr\sigma_1,\, \boldr'\sigma_2,\, \boldr_3\sigma_3,\, \cdots, \boldr_N\sigma_N \right)\right|^2 \label{eq:2nd_order_DM}
\end{align}
This two-particle density may be used to evaluate the expectation value of two-body operators~\cite{Perdew2003DFTprimer}, but it cannot be diagonalized by a unitary transformation of one-electron basis functions~\cite{Davidson1976}. The XC hole density at each coupling strength $\lambda$ is defined as,
\begin{equation}
    n^\lambda_\xc(\boldr, \boldr')= \frac{ n^\lambda_2(\boldr, \boldr') }{ n( \boldr) } -n(\boldr') \label{eq:xcH_lambda_def},
\end{equation}
where the second term removes the classical Hartree contribution to the two-particle density $n( \boldr)n(\boldr')$, with the remaining $\sysXCh(\boldr,\boldr')$ accounting for only the XC effects. The $\lambda$-averaged XC hole density is given by coupling constant integration over this quantity,
\begin{equation}\label{eq:lam_ave_nxc}
    \bar{n}_\xc(\boldr, \boldr')=\int^1_0 \txtd\lambda\, n_\xc^\lambda(\boldr, \boldr'),
\end{equation}
from which an exact expression for $E_\xc$ can be obtained, shown in Eq.~(\ref{eq:Exc_xch_express}). 

At $\lambda=0$, the XC hole is reduces to the exchange hole,
\begin{align}
    n_\xcx(\boldr, \boldr') &= n^{\lambda=0}_\text{xc}(\boldr, \boldr') \notag \\
    &= -\frac{\sum_\sigma \sum^\occ_{i,\,j} \psi^*_{i\sigma}(\boldr) \psi_{j\sigma}(\boldr) \psi^*_{j\sigma}(\boldr') \psi_{i\sigma}( \boldr') }{ n(\boldr)} \label{eq:Xh_def}
\end{align}
where $\psi_{i\sigma}(\boldr)$ are the KS spin-orbitals. Therefore the $\lambda$-averaged correlation hole can be defined by
\begin{equation}
    \bar{n}_\xc( \boldr, \boldr')=n_\xcx\left(\boldr, \boldr'\right)+\bar{n}_{\mathrm{c}}(\boldr, \boldr') \label{eq:Ch_def}.
\end{equation}

Furthermore, since the Coulomb operator has spherical symmetry, the XC energy may be computed \emph{exactly} from the spherically-averaged XC hole. As a result the system- and spherically-averaged XC hole density $\langle\bar{n}_\xc(u)\rangle$ is a useful quantity that can be modelled in order to construct XC energy functionals. This may be written in terms of the distance vector $\boldu=\boldr'-\boldr$ as
\begin{align}
    \langle\bar{n}_\xc(u)\rangle &= \frac{1}{\Ne}\int \dr\, n(\boldr) \int \frac{\txtd\Omega_\boldu }{4\pi} \bar{n}_\xc(\boldr, \boldr+\boldu),
\end{align}
where $\Omega_\boldu$ is the solid angle around direction $\boldu$ and integration is carried out to average over this angle and the spatial coordinates of the entire system. It is a remarkable result that the XC energy may then be expressed precisely as one-dimensional integral over $u = |\boldr'-\boldr|$ for any system,
\begin{align}
    E_\xc\ndep &= \frac{\Ne}{2} \int^\infty_0 \txtd u\, 4\pi u^2 \frac{\langle\bar{n}_\xc(u)\rangle }{u} \label{eq:Exc_def2} \\
    &= \frac{\Ne}{2} \int^\infty_0\du\, \varepsilon_\xc(u)
\end{align}
where we identify $\varepsilon_\xc(u) = 4\pi u\langle \bar{n}_\xc(u) \rangle$. The exact system- and spherically-averaged exchange and correlation holes satisfy the following sum rules respectively,
\begin{align}
    \int^\infty_0\txtd u\, 4\pi u^2 \langle n_\xcx(u) \rangle &=-1, \label{eq:sumrule_X} \\
    \int^\infty_0\txtd u\, 4\pi u^2 \langle\bar{n}_\xcc(u) \rangle &=0. \label{eq:sumrule_C}
\end{align}

\subsection{The Lieb optimization}\label{subsec:lieb_opt}
Given a Hamiltonian $\hat{H}_\lambda[v_\lambda]$, the ground-state energy $E_\lambda[v_\lambda]$ for an $N_{e}$-electron system is given by the Rayleigh-Ritz variation principle as
\begin{equation}\label{eq:rayleigh_ritz}
    E_\lambda[v_\lambda] = \inf_{\Psi_\lambda\in\mathcal{W}_{\Ne}}\left\langle \Psi_\lambda \vert \hat{H}_\lambda[v_\lambda]\vert \Psi_\lambda \right\rangle,
\end{equation}
where $\mathcal{W}_{N_{e}}$ is the set of all $L^{2}$-normalized, antisymmetric $N_{e}$-electron wave functions with a finite kinetic energy. The ground-state energy in Eq.~\eqref{eq:rayleigh_ritz} is well-defined for all potentials $v_\lambda\in\chi^{\ast}$ with $\chi^{\ast} = L^{\frac{3}{2}}+L^{\infty}$, a vector space containing all Coulomb potentials. For a variationally-determined solution to Eq.~\eqref{eq:rayleigh_ritz}, $E_\lambda[v_\lambda]$ is concave and continuous in $v_\lambda$.  

Following the convex-conjugate formulation of DFT by Lieb,~\cite{Lieb1983} the universal density functional $F_\lambda\ndep$ may be defined as the Legendre-Fenchel transform to the ground-state energy of Eq.~\eqref{eq:rayleigh_ritz} as
\begin{equation}\label{eq:lieb_func}
    F_\lambda\ndep = \sup_{v_\lambda \in \chi^{\ast}} \left[ E_\lambda[v_\lambda] - \int \mathrm{d}\mathbf{r}\, n(\mathbf{r})v_\lambda(\mathbf{r}) \right],
\end{equation}
which is convex in $n$ by construction and thus may be defined for arbitrary $E_\lambda[v_\lambda]$. The Legendre-Fenchel transformation of $F_\lambda\ndep$ as defined in Eq.~\eqref{eq:lieb_func} yields an expression for the Hohenberg-Kohn variation principle 
\begin{equation}\label{eq:lieb_biconjugate}
    E_\lambda^{\ast}[v_\lambda] = \inf_{n\in\chi}\left[ F_\lambda\ndep + \int \mathrm{d}\mathbf{r}\, n(\mathbf{r})v_\lambda(\mathbf{r}) \right]
\end{equation}
in which the biconjugate functional $E_\lambda^{\ast}[v_\lambda]$ is the concave envelope to $E_\lambda[v_\lambda]$ such that $E_\lambda^{\ast}[v_\lambda] \geq E_\lambda[v_\lambda]$ and $\chi = L^{3} \cap L^{1}$ is the dual vector space to $\chi^{\ast}$ and which encompasses all $\Ne$-representable densities. The conjugate functionals Eq.~\eqref{eq:lieb_func} and Eq.~\eqref{eq:lieb_biconjugate} are related by Fenchel's inequality as
\begin{equation}\label{eq:fenchel_inequality}
    F_\lambda\ndep \geq E_\lambda^{\ast}[v_\lambda] - \int \mathrm{d}\mathbf{r}\, n(\mathbf{r})v_\lambda(\mathbf{r}) \quad \forall n \in \chi \, , \, v_\lambda \in \chi^{\ast},
\end{equation}
which becomes an equality by maximization of the right-hand side with respect to $v_\lambda$ which is the same, for non-degenerate solutions, as satisfying the stationary condition
\begin{equation}\label{eq:stationary}
    \frac{\delta E_\lambda^{\ast}[v_\lambda]}{\delta v_\lambda(\mathbf{r})} = n(\mathbf{r}).
\end{equation}
By definition, $E_\lambda^{\ast}[v_\lambda]$ is concave in $v_\lambda$ hence has no more than one stationary point; if a solution to Eq.~\eqref{eq:stationary} exists, it is therefore unique. This can also be expressed by re-arrangement of Eq.~\eqref{eq:fenchel_inequality} to the form $E_\lambda^{\ast}[v_\lambda] \leq F_\lambda\ndep + \int \mathrm{d}\mathbf{r}\, n(\mathbf{r})v_\lambda(\mathbf{r})$, which becomes an equality by minimization of the right-hand side with respect to $n(\mathbf{r})$ thus satisfying the stationary condition 
\begin{equation}\label{eq:Fstationary}
    \frac{\delta F_\lambda\ndep}{\delta n(\mathbf{r})} = -v_\lambda(\mathbf{r}),
\end{equation}
where $v_\lambda$ is the optimizing potential. In the Lieb optimization method, the universal density functional $F_\lambda$ is maximized with respect to the potential $v_\lambda(\mathbf{r})$ for a given electronic structure method with energy functional $E_\lambda$ and yielding density $n(\mathbf{r})$. To construct the density-fixed adiabatic connection, the optimizing potential $v_\lambda(\mathbf{r})$ is that for which $E_\lambda$ yields the physically-interacting $\lambda=1$ density for all values of $\lambda\in[0,\,1]$.~\cite{Teale2009,Teale2010}

The universal density functional $F_\lambda$ may be written as a sum of terms according to the Kohn-Sham decomposition as~\cite{KohnSham1965}
\begin{equation}\label{eq:fdecomp}
    F_\lambda\ndep = T_\text{s}\ndep + \lambda E_\text{H}\ndep + \lambda E_\text{x}\ndep + E_{\text{c},\lambda}\ndep,
\end{equation}
in which $T_\text{s}$ is the non-interacting kinetic energy, $E_\text{H}$ is the classical Coulomb energy, $E_\text{x}$ is the exchange energy and $E_{\text{c},\lambda}$ is the $\lambda$-interacting correlation energy. Substituting Eq.~\eqref{eq:fdecomp} into Eq.~\eqref{eq:Fstationary} yields an expression for the optimizing potential in terms of its individual contributions,
\begin{equation}\label{eq:func_deriv_F}
    \begin{aligned}
        \frac{\delta F_\lambda\ndep}{\delta n(\mathbf{r})} &= \frac{\delta T_\text{s}\ndep}{\delta n(\mathbf{r})} + \lambda\frac{\delta E_\text{H}\ndep}{\delta n(\mathbf{r})} + \lambda\frac{\delta E_\text{x}\ndep}{\delta n(\mathbf{r})} + \frac{\delta E_{\text{c},\lambda}\ndep}{\delta n(\mathbf{r})}, \\
        -v_\lambda(\mathbf{r}) &= -v_\text{s}(\mathbf{r}) + \lambda v_\text{H}(\mathbf{r}) + \lambda v_\text{x}(\mathbf{r}) + v_{\text{c},\lambda}(\mathbf{r}).
    \end{aligned}
\end{equation}
Identifying that $v_{\lambda=1} = v_\text{ext}$, the external potential due to the electrostatic potential of the nuclei, and $v_{\lambda=0} = v_\text{s}$, the Kohn-Sham potential may be eliminated from Eq.~\eqref{eq:func_deriv_F} to yield the expression for the optimizing potential at interaction strength $\lambda$ as
\begin{equation}\label{eq:vlambda}
    \begin{aligned}
        v_\lambda(\mathbf{r}) &= v_\text{ext}(\mathbf{r}) + (1-\lambda)v_\text{H}(\mathbf{r}) + (1-\lambda)v_\text{x}(\mathbf{r}) \\
        &+ \lbrack v_{\text{c},1}(\mathbf{r}) - v_{\text{c},\lambda}(\mathbf{r}) \rbrack
    \end{aligned}
\end{equation}
In order to optimize $F_\lambda$ with respect to the potential, it is expanded in a Gaussian basis as proposed by Wu and Yang as~\cite{Yang2002,Wu2003}
\begin{equation}\label{eq:vlambda_wy}
    \begin{aligned}
        v_{\lambda,\mathbf{b}}(\mathbf{r}) &= v_\text{ext}(\mathbf{r}) + (1-\lambda)v_\text{H}(\mathbf{r}) + (1-\lambda)v_\text{ref}(\mathbf{r}) \\
        &+ \sum_{t}b_{t}g_{t}(\mathbf{r}),
    \end{aligned}
\end{equation}
in which $v_\text{H}$ is the Coulomb potential evaluated with an input $\lambda=1$ density $n_\text{in}$, $v_\text{ref}$ is a reference exchange potential also evaluated on this density to ensure that $v_\lambda$ has the correct asymptotic behaviour and $g_{t}$ are a set of Gaussian functions with expansion coefficients $b_{t}$. The form of the reference potential employed in this work is that of the a localized Hartree-Fock potential,~\cite{Sala2001} corrected at long-range by an approximate Fukui potential.~\cite{Parr1984} The details of the construction of the reference potential are given in Appendix~\ref{app:ref_pot}. 

With the parameterization of the potential in Eq.~\eqref{eq:vlambda_wy} the Lieb functional can be defined as an optimization of the objective function
\begin{equation}\label{eq:lieb_obj_fun}
    \mathsf{G}_{\lambda,n}[\mathbf{b}] = E_\lambda[v_{\lambda,\mathbf{b}}] - \int \mathrm{d}\mathbf{r}\, n(\mathbf{r}) v_{\lambda,\mathbf{b}}(\mathbf{r})
\end{equation}
with respect to variations in the potential basis coefficients $\mathbf{b}$; the gradient of Eq.~\eqref{eq:lieb_obj_fun} with respect to the potential basis coefficients is given by
\begin{equation}\label{eq:lieb_obj_grd}
    \frac{\partial \mathsf{G}_{\lambda,n}[\mathbf{b}]}{\partial b_{t}} =  \int \mathrm{d}\mathbf{r}\, \lbrack n_{\lambda,\mathbf{b}}(\mathbf{r})-n_\text{in}(\mathbf{r})\rbrack g_{t}(\mathbf{r})
\end{equation}
whilst the second derivative of the objective function with respect to the potential basis coefficients is given by
\begin{equation}\label{eq:lieb_obj_hess}
    \frac{\partial^{2} \mathsf{G}_{\lambda,n}[\mathbf{b}]}{\partial b_{t}\partial b_{u}} =  \int \int \mathrm{d}\mathbf{r}\,\mathrm{d}\mathbf{r}'\, g_{t}(\mathbf{r}) g_{u}(\mathbf{r}') \frac{\delta n_{\lambda,\mathbf{b}}(\mathbf{r})}{\delta v_{\lambda,\mathbf{b}}(\mathbf{r}')}.
\end{equation}
It can be seen from Eq.~\eqref{eq:lieb_obj_grd} that the stationary condition of Eq.~(\ref{eq:stationary}) will be satisfied where the iterating density $n_{\lambda,\mathbf{b}}$ becomes equal to the input density $n_\text{in}$. In this work, the objective function is optimized by an approximate Newton approach implemented in the \textsc{Quest} code; this is a second-order optimization algorithm in which the Hessian is approximated by the non-interacting Hessian, given by Eq.~\eqref{eq:lieb_obj_hess} at $\lambda=0$.~\cite{Wu2003c} In this process, the potential basis coefficients are updated at each iteration using a backtracking line-search and the wave function $E_\lambda$ evaluated with the corresponding potential $v_{\lambda,\mathbf{b}}$, yielding the energy and iterating density $n_{\lambda,\mathbf{b}}$ from which the objective function Eq.~\eqref{eq:lieb_obj_fun}, gradient Eq.~\eqref{eq:lieb_obj_grd} and approximate Hessian are constructed. At the point of convergence, for which Eq.~\eqref{eq:lieb_obj_grd} becomes zero, the optimizing potential may be used to obtain the $\lambda$-interacting one- and two-particle reduced density matrices required for the construction of the $\lambda$-interacting XC hole as described in Subsection~\ref{subsec:XChole_theo}. With the above calculations completed for each $\lambda$, a series of $\lambda$-dependent and then $\lambda$-averaged quantities such as the XC holes and XC energies given in Eqs.~(\ref{eq:2nd_order_DM}$-$\ref{eq:Exc_def2}) can be readily obtained.

\subsection{The electron-electron cusp condition}\label{subsec:ee_cusp}
For a Coulombic system, the electron-electron cusp condition describes the behavior the electrons in exact eigenfunctions of the Schr{\"o}dinger equation, which exhibit a cusp at the points of electron coalescence due to singularities in the Coulomb potential at such points.~\cite{Kato_cusp/1991PRA} Specifically, the first derivative of the wave function is discontinuous at these points. The electron-electron cusp condition may be expressed using the pair-correlation function, defined as the ratio of the two-particle density to the product of the one-particle densities~\cite{burke1995real}
\begin{equation}\label{eq:pair_corr}
    g(\boldr,\boldr') = \frac{n_2(\boldr,\boldr')}{n(\boldr)n(\boldr')}.
\end{equation}
Given the spherically-averaged pair-correlation functional defined analogously to the spherically-averaged XC hole as $g(\mathbf{r},u) = \int\frac{\txtd\Omega_{\mathbf{u}}}{4\pi}\,g(\mathbf{r},\mathbf{r}+\mathbf{u})$, the electron-electron cusp condition is written as~\cite{kimball1973short,Davidson1976}
\begin{equation}\label{eq:pair_corr_cusp}
    \left.\frac{\partial g(\mathbf{r},u)}{\partial u}\right\vert_{u\rightarrow0^+} = g(\mathbf{r},0)
%    g'(\boldr,\boldr) = g(\boldr,\boldr).
\end{equation}
This may be written in terms of the system and spherically-averaged XC hole defined in Subsection.~\ref{subsec:XChole_theo}, using the relation between $g(\boldr,\boldr')$ and $n_2(\boldr,\boldr')$, as
\begin{equation}\label{eq:xchole_cusp}
    \left.\frac{\partial \langle n_\xc (u)\rangle}{\partial u}\right\vert_{u\rightarrow0^+} = \langle n'_\xc (0)\rangle = \langle n_\xc(0)\rangle +\frac{1}{\Ne} \int\dr\, n^{2}(\boldr).
\end{equation}
Due to the Pauli principle, the cusp condition only arises between electrons with anti-parallel spin and is thus exclusively a correlation effect. The electronic cusp condition can therefore be written in terms of the system and spherically-averaged correlation hole as 
\begin{equation}\label{eq:cuspcond_Chole}
    \left.\frac{\partial \langle n_\xcc (u)\rangle}{\partial u}\right\vert_{u\rightarrow0^+} = \langle n'_\xcc (0)\rangle = \langle n_\xcc(0)\rangle +\frac{1}{2\Ne} \int\dr\, n^{2}(\boldr).
\end{equation}

\subsection{Hookium atoms}\label{subsec:hookium}
A Hookium atom is a model system comprising two electrons confined by a harmonic potential rather than a Coulomb potential,~\cite{taut1993two} with electronic Hamiltonian given in Eq.~(\ref{eq:hookium_orgin}). Introducing the center of mass coordinate $\boldR=(\boldr_1 +\boldr_2)/2$ and the relative separation vector $\boldu=\boldr_1 -\boldr_2$, the Hookium atom Hamiltonian may be resolved into a center of mass and relative separation term as
\begin{equation}\label{eq:hooke_hamiltonian_sep}
    \hat{H} = \underbrace{\left(-\frac{1}{4}\nabla^2_\boldR+k\boldR^2\right)}_{\hat{H}(\mathbf{R})} + \underbrace{\left(-\nabla^2_\boldu +\frac{1}{4}k\boldu^2 +\frac{1}{u}\right)}_{\hat{H}(\mathbf{u})}.
\end{equation}
The second term $\hat{H}(\boldu)$ is of particular interest as it describes the relative motion between the two interacting electrons bound by the harmonic potential and is effectively a one-body problem with Schr{\"o}dinger equation $\hat{H}(\boldu)\varphi(\boldu) = \epsilon\varphi(\boldu)$. Using a separation of variables to write $\varphi(\boldu)$ in terms of the product of radial and angular components
\begin{equation}\label{eq:hookium_phiu}
    \varphi(\boldu) =  \frac{g(u)}{u}  Y_{lm}, \quad g(u) = \exp(-\sqrt{k}u^2/4)T(u),
\end{equation}
where $Y_{lm}$ is the spherical harmonic function describing the angular wave function, a second-order differential equation for $T(u)$ can be obtained. 
By inserting the regular solution
\begin{equation}\label{eq:tu_series}
    T(u) = u \sum^{\infty}_{i=0} a_{i} u^{i}
\end{equation}
into the differential equation, a recurrence relation~\cite{taut1993two} can be found for the coefficients $\{a_i,\enspace i\geq2\}$
\begin{equation}\label{eq:ai_recurrence}
    a_{i+1} = \frac{a_i + \lbrack (i+1/2)\sqrt{k} -\epsilon \rbrack a_{i-1}}{(i+1)(i+2)}
\end{equation}
 where we only consider the ground state with the angular momentum $l=0$. A series of exact solutions can be determined by imposing the condition $a_N=a_{N+1}=0$ at $i=N$, leading to $a_i=0$ for all $i\geq N$. Consequently, $N$ represents the polynomial order of $T(u)$ in the radial wave function $\varphi(u)$. 
 
 For the ground state with $l=0$, $N$ is roughly proportional to $k^{-7.9}$ as observed by fitting the values of $N$ against $k$~\cite{taut1993two}. Since $k$ is the harmonic constant which determines the strength with which electrons are confined, an increase in $N$ implies less confinement and a more radially-diffuse electron density.

However, it is obvious that there doesn't exist an analytical wave function solution for the Hookium atom with the electron-electron interaction scaled by an arbitrary $\lambda \neq 1$. Therefore, the coupling-constant-averaged correlation hole for the Hookium atom has seldom been studied, and only the correlation hole at $\lambda=1$ has been comprehensively studied~\cite{qian1998physics} and used to benchmark correlation hole models~\cite{constantin2013construction, burke1994validity, burke1994local}.

For example, the exchange hole and the correlation hole of the Hookium atom with $k=1/4$ (corresponding to $N=2$) for the $\lambda=1$ case have been carefully studied in Ref.~\citenum{qian1998physics}, which is also used to benchmark the system- and angle-averaged XC hole models of different meta-GGAs~\cite{constantin2013construction}. Using only the $\lambda=1$ results, the validity of the electronic cusp condition in the ground state of the Hookium atom for arbitrary harmonic confinement $k$ has been demonstrated ~\cite{burke1994validity} and it has been demonstrated that the LDA hole model can precisely capture the cusp condition of the Hookium atom~\cite{burke1994local}. 

In this work, we employ the exact solutions of the Hookium atom at $\lambda=1$ to benchmark those calculated from the CCSD wave function. We then use the Lieb optimization with a CCSD wave function to calculate $\lambda$-averaged XC holes for the Hookium atom, which can serve as a benchmark for XC hole models.

\section{Computational details}\label{sec:computation}
In this work all calculations are carried out using the \textsc{Quest} code with a the spin-restricted CCSD wave function as the reference method. The convergence of self-consistent field calculations was accelerated using the C1-DIIS method, with a convergence threshold of $10^{-12}$ a.u. on the norm of the DIIS error vector. For the CCSD calculations the convergence threshold for both the excitation amplitudes and de-excitation amplitudes, required for evaluation of the CCSD one- and two-particle densities, was $10^{-10}$ a.u. for the norm of the difference of the amplitudes between iterations. 

Lieb optimizations were carried out with at the CCSD level for a range of interaction strengths $\lambda\in[0,\,1]$ using the approximate Netwon method described in Subsection~\ref{subsec:lieb_opt}. In each case the CCSD $\lambda=1$ density was used as input to the Lieb functional, in order to fix the density along the adibatic connection at its physical value. The potential was regularised using the smoothing norm method with a regularization parameter of $10^{-5}$ a.u. Convergence of the Lieb optimization was based on the norm of the gradient with respect to potential expansion coefficients, with a convergence threshold of $10^{-8}$ a.u. used throughout. To improve convergence, a slightly smaller basis set was used for the potential expansion than was used for the orbital expansion: in this work a series of Dunning basis sets were employed, with the orbital basis sets $Y$-aug-cc-pV\textit{X}Z ($Y=$d, t, q, 5, 6; $X=$D, T, Q, 5, 6) and corresponding potential basis sets of ($Y-1$)-aug-cc-pV\textit{X}Z ($Y-1=$d, t, q, 5; $X=$D, T, Q, 5, 6).~\cite{Dunning1989,Woon1993,Woon1995} In each case, the basis sets were uncontracted spherical Gaussians with exponents for the Helium atom used throughout. 

To evaluate the system- and spherically-averaged XC holes, a nested numerical quadrature was employed. The spherically-averaged XC hole $n_\text{xc}^\lambda$ was constructed by angular integration using an order-41 Lebedev quadrature grid at each reference point~\cite{Lebedev1976,Lebedev1992}, leading to,
\begin{equation}\label{eq:sph_ave_quad}
    \begin{aligned}
        n_\text{xc}^\lambda(\mathbf{r},u) &= \frac{1}{4\pi}\int\mathrm{d}\Omega_{\mathbf{u}} n_\text{xc}^\lambda(\mathbf{r},\mathbf{r}+\mathbf{u}) \\
        &\approx \sum_{i}^{N_{\Omega}} w_{i}^{\Omega} n_\text{xc}^\lambda(\mathbf{r},\mathbf{r}_{i}^{\Omega}),
    \end{aligned}
\end{equation}
with quadrature weights $w_{i}^{\Omega}$ and nodes $\mathbf{r}_{i}^{\Omega}$ associated to the angular quadrature nodes $(\varphi_{i},\theta_{i})$ by
\begin{equation*}
    \begin{aligned}
        x_{i}^{\Omega} &= u \cos{\varphi_{i}} \sin{\theta_{i}} + x \\
        y_{i}^{\Omega} &= u \sin{\varphi_{i}} \sin{\theta_{i}} + y \\
        z_{i}^{\Omega} &= u \cos{\theta_{i}} + z,
    \end{aligned}
\end{equation*}
with $\mathbf{r} = (x,y,z)$ and $\mathbf{r}_{i}^{\Omega} = (x_{i}^{\Omega},y_{i}^{\Omega},z_{i}^{\Omega})$. The system-averaging  was then carried-out by numerical integration of the reference point using a full quadrature grid, with angular component again given by the order-41 Lebedev quadrature and radial component constructed using the scheme of Lindh, Malmqvist and Gagliardi~\cite{Lindh2001} with a relative error threshold of $10^{-10}$ a.u.,
\begin{equation}\label{eq:sys_ave_quad}
    \begin{aligned}
        \langle n_\text{xc}^\lambda(u) \rangle &= \frac{1}{\Ne}\int\mathrm{d}\mathbf{r}\, n(\mathbf{r}) n_\text{xc}^\lambda(\mathbf{r},u) \\
        &\approx \sum_{i}^{N_{r}}\sum_{j}^{N_{\Omega}} w_{i}^{r} w_{j}^{\Omega} n(\mathbf{r}_{ij}) n_\text{xc}^\lambda(\mathbf{r}_{ij},u),
    \end{aligned}
\end{equation}
where $w_{i}^{r}$ are the weights of the radial quadrature and $\mathbf{r}_{ij}$ the product of radial quadrature nodes $\mathbf{r}_{i}^{r}$ and angular quadrature nodes $\mathbf{r}_{j}^{\Omega}$. 

Exact analytical results for the Hookium atom at $\lambda=1$, as described in Subsection~\ref{subsec:hookium}, were also calculated with Mathematica, allowing us to carefully assess the accuracy of the finite basis CCSD calculations. 

\section{Results and discussion}
\label{sec:results_discuss}
\subsection{Accuracy of finite-basis CCSD Hookium solutions}\label{subsec:ccsd_totene_density}

As described in Subsection.~\ref{subsec:lieb_opt}, $n(\boldr)$ given by CCSD is used as the reference electron density of the physical interacting system for the Lieb optimization in Eq.~\eqref{eq:lieb_obj_fun}. Therefore, we first assess the quality of CCSD calculated total energies and densities with a range of orbital basis set sizes for Hookium by comparing them with the exact analytical results.% computed as described in Subsection~\ref{subsec:hookium}. 
 
\subsubsection{Total energies}\label{subsubsec:ccsd_tote}
\begin{table*}[!htp]\centering\renewcommand{\arraystretch}{1.24}
\setlength{\tabcolsep}{0pt}
\begin{threeparttable}
\caption{Percentage errors of the CCSD total energy $E^\ccsd_\text{tot}$ relative to the exact results for the Hookium atom solutions with $N=2\sim11$. Orbital basis sets of $Y$-aug-cc-pV\textit{X}Z ($X=$D, T, Q, 5, 6; $Y=$d, t, q, p, s) are employed.}
\label{tab:PEs_Etot_CCSD}
\begin{tabular*}{.72\linewidth}{@{\extracolsep{\fill}}lzzzzzzzzzzz}
\toprule
\multicolumn{1}{c}{Orbital basis} &
\multicolumn{1}{c}{\,$N=2$\,} &
  \multicolumn{1}{c}{\,$N=3$\,} &
  \multicolumn{1}{c}{$\,N=4\,$} &
  \multicolumn{1}{c}{$N=5$} &
  \multicolumn{1}{c}{$N=6$} &
  \multicolumn{1}{c}{$N=7$} &
  \multicolumn{1}{c}{$N=8$} &
  \multicolumn{1}{c}{$N=9$} &
  \multicolumn{1}{c}{$N=10$} &
  \multicolumn{1}{c}{$N=11$} \\ \midrule
d-aug-cc-pVDZ &
  \cellcolor[HTML]{B0D47F}1.9\% &
  \cellcolor[HTML]{FFEB84}4.5\% &
  \cellcolor[HTML]{FFEB84}8.6\% &
  \cellcolor[HTML]{FFE483}53.6\% &
  \cellcolor[HTML]{FED881}129.1\% &
  \cellcolor[HTML]{FEC87E}228.1\% &
  \cellcolor[HTML]{FDB57A}348.5\% &
  \cellcolor[HTML]{FB9F76}489.6\% &
  \cellcolor[HTML]{FA8671}651.0\% &
  \cellcolor[HTML]{F8696B}832.6\% \\
d-aug-cc-pVTZ &
  \cellcolor[HTML]{63BE7B}0.2\% &
  \cellcolor[HTML]{9ACD7E}1.4\% &
  \cellcolor[HTML]{FFEB84}6.5\% &
  \cellcolor[HTML]{FFE784}34.9\% &
  \cellcolor[HTML]{FFDD82}93.5\% &
  \cellcolor[HTML]{FED17F}173.8\% &
  \cellcolor[HTML]{FDC17C}272.6\% &
  \cellcolor[HTML]{FCAF79}388.9\% &
  \cellcolor[HTML]{FB9A75}522.3\% &
  \cellcolor[HTML]{FA8370}672.5\% \\
d-aug-cc-pVQZ &
  \cellcolor[HTML]{6CC07B}0.4\% &
  \cellcolor[HTML]{7FC67C}0.8\% &
  \cellcolor[HTML]{FFEB84}5.6\% &
  \cellcolor[HTML]{FFE784}33.0\% &
  \cellcolor[HTML]{FFDE82}90.2\% &
  \cellcolor[HTML]{FED280}168.8\% &
  \cellcolor[HTML]{FDC27D}265.7\% &
  \cellcolor[HTML]{FCB179}379.7\% &
  \cellcolor[HTML]{FB9C75}510.5\% &
  \cellcolor[HTML]{FA8571}657.9\% \\
d-aug-cc-pV5Z &
  \cellcolor[HTML]{71C27B}0.5\% &
  \cellcolor[HTML]{75C37C}0.6\% &
  \cellcolor[HTML]{FFEB84}4.7\% &
  \cellcolor[HTML]{FFE784}32.7\% &
  \cellcolor[HTML]{FFDE82}90.1\% &
  \cellcolor[HTML]{FED280}168.8\% &
  \cellcolor[HTML]{FDC27D}265.7\% &
  \cellcolor[HTML]{FCB179}379.8\% &
  \cellcolor[HTML]{FB9C75}510.7\% &
  \cellcolor[HTML]{FA8571}658.1\% \\
d-aug-cc-pV6Z &
  \cellcolor[HTML]{68BF7B}0.3\% &
  \cellcolor[HTML]{66BE7B}0.3\% &
  \cellcolor[HTML]{FFEB84}4.6\% &
  \cellcolor[HTML]{FFE784}34.5\% &
  \cellcolor[HTML]{FFDD82}93.6\% &
  \cellcolor[HTML]{FED17F}174.2\% &
  \cellcolor[HTML]{FDC17C}273.3\% &
  \cellcolor[HTML]{FCAF79}389.8\% &
  \cellcolor[HTML]{FB9A75}523.5\% &
  \cellcolor[HTML]{FA8270}674.0\% \\ \hline
t-aug-cc-pVDZ &
  \cellcolor[HTML]{AFD37F}1.9\% &
  \cellcolor[HTML]{FFEB84}3.8\% &
  \cellcolor[HTML]{FFEB84}4.2\% &
  \cellcolor[HTML]{B4D57F}2.0\% &
  \cellcolor[HTML]{FFEA84}11.9\% &
  \cellcolor[HTML]{FFE683}40.2\% &
  \cellcolor[HTML]{FFDF82}81.8\% &
  \cellcolor[HTML]{FED781}133.8\% &
  \cellcolor[HTML]{FECE7F}194.9\% &
  \cellcolor[HTML]{FDC37D}264.5\% \\
t-aug-cc-pVTZ &
  \cellcolor[HTML]{63BE7B}0.2\% &
  \cellcolor[HTML]{95CC7D}1.3\% &
  \cellcolor[HTML]{74C37C}0.6\% &
  \cellcolor[HTML]{FFEB84}6.5\% &
  \cellcolor[HTML]{FFEA84}14.0\% &
  \cellcolor[HTML]{FFE784}35.3\% &
  \cellcolor[HTML]{FFE182}70.4\% &
  \cellcolor[HTML]{FFDA81}116.2\% &
  \cellcolor[HTML]{FED17F}170.9\% &
  \cellcolor[HTML]{FEC77E}233.8\% \\
t-aug-cc-pVQZ &
  \cellcolor[HTML]{6BC07B}0.4\% &
  \cellcolor[HTML]{7CC57C}0.8\% &
  \cellcolor[HTML]{6FC17B}0.5\% &
  \cellcolor[HTML]{FFEB84}5.7\% &
  \cellcolor[HTML]{FFEA84}14.1\% &
  \cellcolor[HTML]{FFE684}36.8\% &
  \cellcolor[HTML]{FFE182}73.2\% &
  \cellcolor[HTML]{FFD981}120.2\% &
  \cellcolor[HTML]{FED07F}176.3\% &
  \cellcolor[HTML]{FDC67D}240.6\% \\
t-aug-cc-pV5Z &
  \cellcolor[HTML]{70C17B}0.5\% &
  \cellcolor[HTML]{71C27B}0.5\% &
  \cellcolor[HTML]{72C27B}0.5\% &
  \cellcolor[HTML]{FFEB84}4.3\% &
  \cellcolor[HTML]{FFEA84}14.0\% &
  \cellcolor[HTML]{FFE683}39.3\% &
  \cellcolor[HTML]{FFE082}78.1\% &
  \cellcolor[HTML]{FED881}127.5\% &
  \cellcolor[HTML]{FECF7F}186.0\% &
  \cellcolor[HTML]{FDC47D}253.0\% \\
t-aug-cc-pV6Z &
  \cellcolor[HTML]{67BF7B}0.3\% &
  \cellcolor[HTML]{63BE7B}0.2\% &
  \cellcolor[HTML]{6CC07B}0.4\% &
  \cellcolor[HTML]{FFEB84}3.8\% &
  \cellcolor[HTML]{FFE984}17.5\% &
  \cellcolor[HTML]{FFE583}48.3\% &
  \cellcolor[HTML]{FFDE82}92.6\% &
  \cellcolor[HTML]{FED580}147.7\% &
  \cellcolor[HTML]{FECB7E}212.5\% &
  \cellcolor[HTML]{FDBF7C}286.3\% \\ \hline
q-aug-cc-pVDZ &
  \cellcolor[HTML]{AED37F}1.9\% &
  \cellcolor[HTML]{FFEB84}3.7\% &
  \cellcolor[HTML]{FFEB84}3.9\% &
  \cellcolor[HTML]{9BCE7E}1.5\% &
  \cellcolor[HTML]{FFEB84}6.0\% &
  \cellcolor[HTML]{EBE582}3.3\% &
  \cellcolor[HTML]{F3E783}3.5\% &
  \cellcolor[HTML]{FFEA84}13.4\% &
  \cellcolor[HTML]{FFE784}31.1\% &
  \cellcolor[HTML]{FFE483}54.7\% \\
q-aug-cc-pVTZ &
  \cellcolor[HTML]{63BE7B}0.2\% &
  \cellcolor[HTML]{94CC7D}1.3\% &
  \cellcolor[HTML]{70C17B}0.5\% &
  \cellcolor[HTML]{FBEA83}3.6\% &
  \cellcolor[HTML]{7BC57C}0.7\% &
  \cellcolor[HTML]{FFEB84}6.6\% &
  \cellcolor[HTML]{FFEA84}14.0\% &
  \cellcolor[HTML]{FFE884}24.8\% &
  \cellcolor[HTML]{FFE683}41.5\% &
  \cellcolor[HTML]{FFE283}64.5\% \\
q-aug-cc-pVQZ &
  \cellcolor[HTML]{6BC07B}0.4\% &
  \cellcolor[HTML]{7BC57C}0.7\% &
  \cellcolor[HTML]{6AC07B}0.4\% &
  \cellcolor[HTML]{D4DE81}2.8\% &
  \cellcolor[HTML]{82C77C}0.9\% &
  \cellcolor[HTML]{FFEB84}7.2\% &
  \cellcolor[HTML]{FFEA84}15.1\% &
  \cellcolor[HTML]{FFE884}27.5\% &
  \cellcolor[HTML]{FFE583}46.5\% &
  \cellcolor[HTML]{FFE182}71.9\% \\
q-aug-cc-pV5Z &
  \cellcolor[HTML]{70C17B}0.5\% &
  \cellcolor[HTML]{70C17B}0.5\% &
  \cellcolor[HTML]{6DC07B}0.4\% &
  \cellcolor[HTML]{AED37F}1.9\% &
  \cellcolor[HTML]{98CD7E}1.4\% &
  \cellcolor[HTML]{FFEB84}6.2\% &
  \cellcolor[HTML]{FFEA84}14.7\% &
  \cellcolor[HTML]{FFE784}30.0\% &
  \cellcolor[HTML]{FFE483}52.4\% &
  \cellcolor[HTML]{FFDF82}81.0\% \\
q-aug-cc-pV6Z &
  \cellcolor[HTML]{67BF7B}0.3\% &
  \cellcolor[HTML]{63BE7B}0.2\% &
  \cellcolor[HTML]{68BF7B}0.3\% &
  \cellcolor[HTML]{7EC67C}0.8\% &
  \cellcolor[HTML]{A7D17E}1.7\% &
  \cellcolor[HTML]{FFEB84}7.1\% &
  \cellcolor[HTML]{FFE984}20.0\% &
  \cellcolor[HTML]{FFE683}41.5\% &
  \cellcolor[HTML]{FFE182}70.5\% &
  \cellcolor[HTML]{FFDC81}105.7\% \\ \hline
p-aug-cc-pVDZ &
  \cellcolor[HTML]{AED37F}1.9\% &
  \cellcolor[HTML]{FFEB84}3.7\% &
  \cellcolor[HTML]{FFEB84}3.9\% &
  \cellcolor[HTML]{98CD7E}1.4\% &
  \cellcolor[HTML]{FFEB84}5.7\% &
  \cellcolor[HTML]{D3DE81}2.7\% &
  \cellcolor[HTML]{DDE182}3.0\% &
  \cellcolor[HTML]{FFEB84}7.8\% &
  \cellcolor[HTML]{FFEB84}6.1\% &
  \cellcolor[HTML]{FBE983}3.6\% \\
p-aug-cc-pVTZ &
  \cellcolor[HTML]{63BE7B}0.2\% &
  \cellcolor[HTML]{94CC7D}1.3\% &
  \cellcolor[HTML]{6FC17B}0.5\% &
  \cellcolor[HTML]{F2E783}3.4\% &
  \cellcolor[HTML]{77C37C}0.7\% &
  \cellcolor[HTML]{FFEB84}4.2\% &
  \cellcolor[HTML]{A6D17E}1.7\% &
  \cellcolor[HTML]{92CB7D}1.3\% &
  \cellcolor[HTML]{FFEB84}6.8\% &
  \cellcolor[HTML]{FFEA84}14.4\% \\
p-aug-cc-pVQZ &
  \cellcolor[HTML]{6BC07B}0.4\% &
  \cellcolor[HTML]{7BC57C}0.7\% &
  \cellcolor[HTML]{6AC07B}0.3\% &
  \cellcolor[HTML]{CADB80}2.5\% &
  \cellcolor[HTML]{7BC47C}0.7\% &
  \cellcolor[HTML]{E8E482}3.2\% &
  \cellcolor[HTML]{85C77C}1.0\% &
  \cellcolor[HTML]{EAE482}3.3\% &
  \cellcolor[HTML]{FFEA84}10.3\% &
  \cellcolor[HTML]{FFE984}18.1\% \\
p-aug-cc-pV5Z &
  \cellcolor[HTML]{70C17B}0.5\% &
  \cellcolor[HTML]{70C17B}0.5\% &
  \cellcolor[HTML]{6CC07B}0.4\% &
  \cellcolor[HTML]{A4D07E}1.7\% &
  \cellcolor[HTML]{89C97D}1.1\% &
  \cellcolor[HTML]{BED880}2.3\% &
  \cellcolor[HTML]{93CB7D}1.3\% &
  \cellcolor[HTML]{FFEB84}4.4\% &
  \cellcolor[HTML]{FFEA84}10.3\% &
  \cellcolor[HTML]{FFE984}18.6\% \\
p-aug-cc-pV6Z &
  \cellcolor[HTML]{67BF7B}0.3\% &
  \cellcolor[HTML]{63BE7B}0.2\% &
  \cellcolor[HTML]{67BF7B}0.3\% &
  \cellcolor[HTML]{76C37C}0.6\% &
  \cellcolor[HTML]{90CB7D}1.2\% &
  \cellcolor[HTML]{83C77C}0.9\% &
  \cellcolor[HTML]{A5D17E}1.7\% &
  \cellcolor[HTML]{FFEB84}6.0\% &
  \cellcolor[HTML]{FFEA84}14.4\% &
  \cellcolor[HTML]{FFE884}27.2\% \\ \hline
s-aug-cc-pVDZ &
  \cellcolor[HTML]{AED37F}1.9\% &
  \cellcolor[HTML]{FEEA83}3.7\% &
  \cellcolor[HTML]{FFEB84}3.9\% &
  \cellcolor[HTML]{98CD7E}1.4\% &
  \cellcolor[HTML]{FFEB84}5.6\% &
  \cellcolor[HTML]{D0DD81}2.7\% &
  \cellcolor[HTML]{DCE182}2.9\% &
  \cellcolor[HTML]{FFEB84}7.3\% &
  \cellcolor[HTML]{FFEB84}5.6\% &
  \cellcolor[HTML]{F0E683}3.4\% \\
s-aug-cc-pVTZ &
  \cellcolor[HTML]{63BE7B}0.2\% &
  \cellcolor[HTML]{94CC7D}1.3\% &
  \cellcolor[HTML]{6FC17B}0.5\% &
  \cellcolor[HTML]{F0E683}3.4\% &
  \cellcolor[HTML]{77C37C}0.6\% &
  \cellcolor[HTML]{FFEB84}4.0\% &
  \cellcolor[HTML]{9BCE7E}1.5\% &
  \cellcolor[HTML]{8FCA7D}1.2\% &
  \cellcolor[HTML]{FFEB84}4.5\% &
  \cellcolor[HTML]{F5E883}3.5\% \\
s-aug-cc-pVQZ &
  \cellcolor[HTML]{6BC07B}0.4\% &
  \cellcolor[HTML]{7BC57C}0.7\% &
  \cellcolor[HTML]{6AC07B}0.3\% &
  \cellcolor[HTML]{C8DB80}2.5\% &
  \cellcolor[HTML]{7AC47C}0.7\% &
  \cellcolor[HTML]{DCE182}2.9\% &
  \cellcolor[HTML]{77C47C}0.7\% &
  \cellcolor[HTML]{B9D780}2.2\% &
  \cellcolor[HTML]{EBE582}3.3\% &
  \cellcolor[HTML]{ACD37F}1.8\% \\
s-aug-cc-pV5Z &
  \cellcolor[HTML]{70C17B}0.5\% &
  \cellcolor[HTML]{70C17B}0.5\% &
  \cellcolor[HTML]{6CC07B}0.4\% &
  \cellcolor[HTML]{A3D07E}1.6\% &
  \cellcolor[HTML]{88C87D}1.0\% &
  \cellcolor[HTML]{B2D47F}2.0\% &
  \cellcolor[HTML]{79C47C}0.7\% &
  \cellcolor[HTML]{C2D980}2.4\% &
  \cellcolor[HTML]{CADB80}2.5\% &
  \cellcolor[HTML]{B6D67F}2.1\% \\
s-aug-cc-pV6Z &
  \cellcolor[HTML]{67BF7B}0.3\% &
  \cellcolor[HTML]{63BE7B}0.2\% &
  \cellcolor[HTML]{67BF7B}0.3\% &
  \cellcolor[HTML]{75C37C}0.6\% &
  \cellcolor[HTML]{8CCA7D}1.1\% &
  \cellcolor[HTML]{77C37C}0.6\% &
  \cellcolor[HTML]{90CB7D}1.2\% &
  \cellcolor[HTML]{A1D07E}1.6\% &
  \cellcolor[HTML]{86C87D}1.0\% &
  \cellcolor[HTML]{C9DB80}2.5\% \\
\midrule
\bottomrule
\end{tabular*}
%\begin{tablenotes}
%\item[a]\label{expt_vals} Experiment
%\end{tablenotes}
\end{threeparttable}
\end{table*}

Table~\ref{tab:PEs_Etot_CCSD} lists the percentage errors (PEs) for the finite-basis CCSD total energies with respect to the exact results for different solutions to the Hookium atom labelled by $N$, as described in Section~\ref{subsec:hookium}, computed with different basis sets. All PEs are positive, as expected since CCSD is equivalent to FCI for these two-electron systems, and so the energy approaches the complete basis FCI energy from above. In general, the accuracy of the energies can be improved systematically by using basis sets with a higher cardinal number \textit{X} or higher augmentation with diffuse functions $Y$. This leads to a reduction in the PEs to be in the range $0.2\%$ -- $2.5\%$. It is clear that for the solutions with $N<5$ PEs below 1\% can be achieved with triply augmented basis sets with cardinal numbers of $4$ or above. Indeed, adding extra diffuse functions does not further improve the accuracy of the results for these solutions. However, for larger values of $N$ is it is essential to include many more diffuse functions to obtain reasonable accuracy. For $5 \leq N \leq 8$ pentuple augmentation is required to achieve PEs below $2\%$ and for $N>8$ hextuple augmentation is required. The dependence on cardinal number \textit{X} is less significant, once sufficient diffuse functions are included for a given value of $N$, there appears to be little benefit in using basis sets with $X>4$.

%This leads to a reduction in the PE to less than $1.5\%$, with the exception of the $N=11$ solution, for which the charge density is highly diffuse. For a given basis set, the lower the order $N$ of the solution, the greater the accuracy achieved as expected. For lower $N$ solutions (e.g. $N=2$), with more compact electron densities, the inclusion of more diffuse functions in the basis set doesn't further improve the accuracy of the results.

%We also note that, with the same number of diffuse functions (i.e. the same value of Y in the Y-aug-cc-pVXZ basis set series), calculations using double-$\zeta$ basis always give significant PEs, indicating that such basis sets are usually insufficient for the Hookium atom. This can be observed by examining the PEs of Y-aug-cc-pVDZ (Y$=$d, t, q, 5, 6) for solutions of different $N$ values in Table~\ref{tab:PEs_Etot_CCSD}.

\subsubsection{Electron densities}\label{subsubsec:ccsd_dens}
\begin{figure*}[!htp]\centering
\includegraphics[width=.6\textwidth]{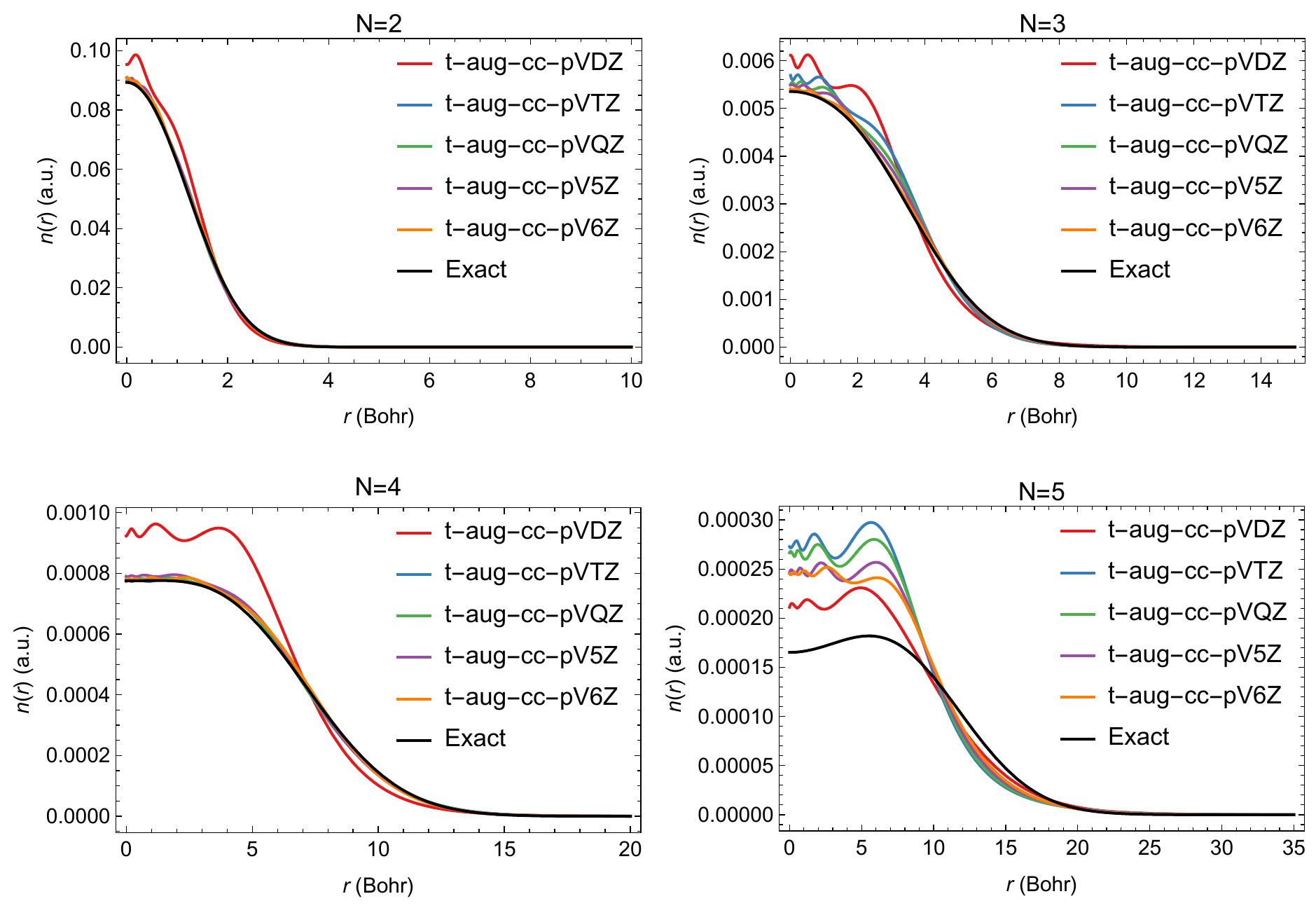}
\caption{Radical electron densities calculated from CCSD for Hookium solutions with $N=2,\, 3,\, 4,\, 5$, where orbital basis sets with t-aug-cc-pV\textit{X}Z ($X=$D, T, Q, 5, 6) are employed.}
\label{fig:ccsd_density}
\end{figure*}

\begin{table}[!htp]\centering\renewcommand{\arraystretch}{1.24}
\setlength{\tabcolsep}{0pt}
\begin{threeparttable}
\caption{Absolute percentage errors of CCSD electron densities estimated by Eq.~\eqref{eq:ccsd_density_PE} for the Hookium solutions with $N=2\sim5$ in the t-aug-cc-pV\textit{X}Z ($X=$ D, T, Q, 5, 6) basis sets.}
\label{tab:PEs_density_CCSD}
\begin{tabular*}{\linewidth}{@{\extracolsep{\fill}}lzzzz}
\toprule
\multicolumn{1}{c}{Orbital basis}&\multicolumn{1}{c}{$N=2$} &\multicolumn{1}{c}{$ N=3$} &\multicolumn{1}{c}{$N=4$} &\multicolumn{1}{c}{$N=5$} \\
\midrule
%t-aug-cc-pVDZ & \cellcolor[HTML]{99CF86}13.10\% & \cellcolor[HTML]{AED58B}17.71\% & \cellcolor[HTML]{BFDB8E}21.28\% & \cellcolor[HTML]{A0D187}14.63\% \\
%t-aug-cc-pVTZ & \cellcolor[HTML]{63BE7B}1.59\%  & \cellcolor[HTML]{96CE85}12.57\% & \cellcolor[HTML]{6CC07C}3.51\%  & \cellcolor[HTML]{FFEF9C}34.73\% \\
%t-aug-cc-pVQZ &  \cellcolor[HTML]{6DC17D}3.79\% &  \cellcolor[HTML]{83C881}8.56\% &  \cellcolor[HTML]{6BC07C}3.31\% &  \cellcolor[HTML]{F0EA98}31.55\% \\
%t-aug-cc-pV5Z &  \cellcolor[HTML]{6FC17D}4.27\% &  \cellcolor[HTML]{77C47F}5.99\% &  \cellcolor[HTML]{6EC17D}3.97\% &  \cellcolor[HTML]{D4E193}25.76\% \\
%t-aug-cc-pV6Z &  \cellcolor[HTML]{6AC07C}3.10\% &  \cellcolor[HTML]{68BF7C}2.76\% &  \cellcolor[HTML]{6DC17D}3.74\% &  \cellcolor[HTML]{BCDA8E}20.67\% \\
t-aug-cc-pVDZ  &13.10\%  &17.71\%  &21.28\%  &14.63\% \\
t-aug-cc-pVTZ  & 1.59\%  &12.57\%  & 3.51\%  &34.73\% \\
t-aug-cc-pVQZ  & 3.79\%  & 8.56\%  & 3.31\%  &31.55\% \\
t-aug-cc-pV5Z  & 4.27\%  & 5.99\%  & 3.97\%  &25.76\% \\
t-aug-cc-pV6Z  & 3.10\%  & 2.76\%  & 3.74\%  &20.67\% \\
\midrule
\bottomrule
\end{tabular*}
%\begin{tablenotes}
%\item[a]\label{FeO_expt} Ref.~\cite{Laisheng1996/JAmChemSoc.118.5296}.
%\end{tablenotes}
\end{threeparttable}
\end{table}
In Figure~\ref{fig:ccsd_density}, we plot the CCSD electron densities of Hookium atom solutions with $2 \leq N \leq 5$ radially from the atomic nucleus. For comparison the densities of the corresponding exact solutions are also shown. The convergence of CCSD electron densities at each Hookium solution $N$ is examined by gradually increasing the size of the t-aug-cc-pV\textit{X}Z basis set by changing the cardinal number \textit{X} from $2$ to $6$.

Figure~\ref{fig:ccsd_density} shows that, as the order of the Hookium solution $N$ increases from $2$ to $5$, the corresponding electron density becomes increasingly spatially diffuse.
Interestingly, the convergence behavior of the density with respect to basis set appears to be dependent on whether the value of $N$ is even or odd. Specifically, for the $N=2$ and $N=4$ solutions, the CCSD densities converge to the corresponding exact densities relatively quickly with increasing basis set size, and there is no discernible difference in the results obtained with a basis sets with $X \geq 3$. 
However, convergence of the density with respect to the basis set is considerably slower for the $N=3$ and $N=5$ solutions. It should be noted that, for $N=5$, the CCSD energies and densities in the largest basis set t-aug-cc-pV6Z both have a greater error than those from the smallest basis set t-aug-cc-pVDZ considered here, as can be seen in Table~\ref{tab:PEs_Etot_CCSD}.% and Table~\ref{tab:PEs_density_CCSD}.
This indicates that for $N=5$ the triply augmented basis sets are not adequately diffuse and errors could be reduced by further augmentation of the basis set. Indeed, from the analysis of the electron density it is clear that a sufficiently diffuse basis set would be required to to represent the electron density accurately as $N$ increases, consistent with the analysis of the CCSD total energies in Section~\ref{subsubsec:ccsd_tote}. 

To quantify the deviations of the CCSD electron densities from those of the exact solutions in Figure~\ref{fig:ccsd_density}, the absolute percentage error ($\aPE$) is defined as
\begin{align}
    \aPE &= \frac{1}{2} \int\dr \left|n^\ccsd(\boldr) -n^\exact(\boldr)\right| \times 100\%
\label{eq:ccsd_density_PE}
\end{align}
where the division by $2$, the number of electrons in the Hookium atom, is to give the absolute percentage error per electron. The results are presented in Table~\ref{tab:PEs_density_CCSD}, showing that for basis sets with $X>2$, the PEs for $N=2$ and $N=4$ are consistently lower than $4\%$. With the t-aug-cc-pV6Z basis set, CCSD calculations yield less than $4\%$ PEs for $N=3$. However, for $N=5$ the PEs are greater than $20\%$ for all t-aug-cc-pV\textit{X}Z basis sets, with the exception of the t-aug-cc-pVDZ basis set. Therefore, we only consider Hookium solutions with $N=2,\,3,\,4$ in the subsequent XC hole calculations using CCSD+Lib.
 
\subsection{Accuracy of finite-basis CCSD XC holes at $\lambda=1$}\label{subsec:XChole_theos_lambda1_basiseffects}
We now employ the Lieb optimization at the CCSD level to compute the exchange holes at $\lambda = 0$, correlation holes and XC holes at $\lambda=1$, for Hookium solutions with $N=2,\,3,\,4$. %as the difference between these functions. The Hookium atom solutions with $N=2,\,3,\,4$ are compared with those obtained from the exact solutions.
The main focus of this analysis is to determine the effect of the electron-electron cusp, and the limitations of finite Gaussian basis sets in its representation%,
on the correlation hole.% However, 
Moreover, we also consider the effect of basis set size on the exchange energy $\Ex$ and correlation energy $\Ec^{\lambda=1}$ for the Hookium atom solutions with $2 \leq N \leq 11$. 

When comparing the CCSD $\lambda=1$ XC holes with those of the exact solutions, the errors are dominated by incompleteness of the finite basis set in which the orbitals are expanded. However, when comparing the CCSD $\lambda=0$ exchange and $\lambda=1$ correlation holes with those of the exact solutions, an additional source of error is introduced; the incompleteness of the basis set in which the potential is expanded, shown in Eq.~\eqref{eq:vlambda_wy}, and the associated numerical errors in the convergence of the Lieb optimization at $\lambda=0$. 

\subsubsection{The exchange hole}\label{subsubsec:exchange_hole}

\begin{figure*}[!htp]\centering
\includegraphics[width=\textwidth]{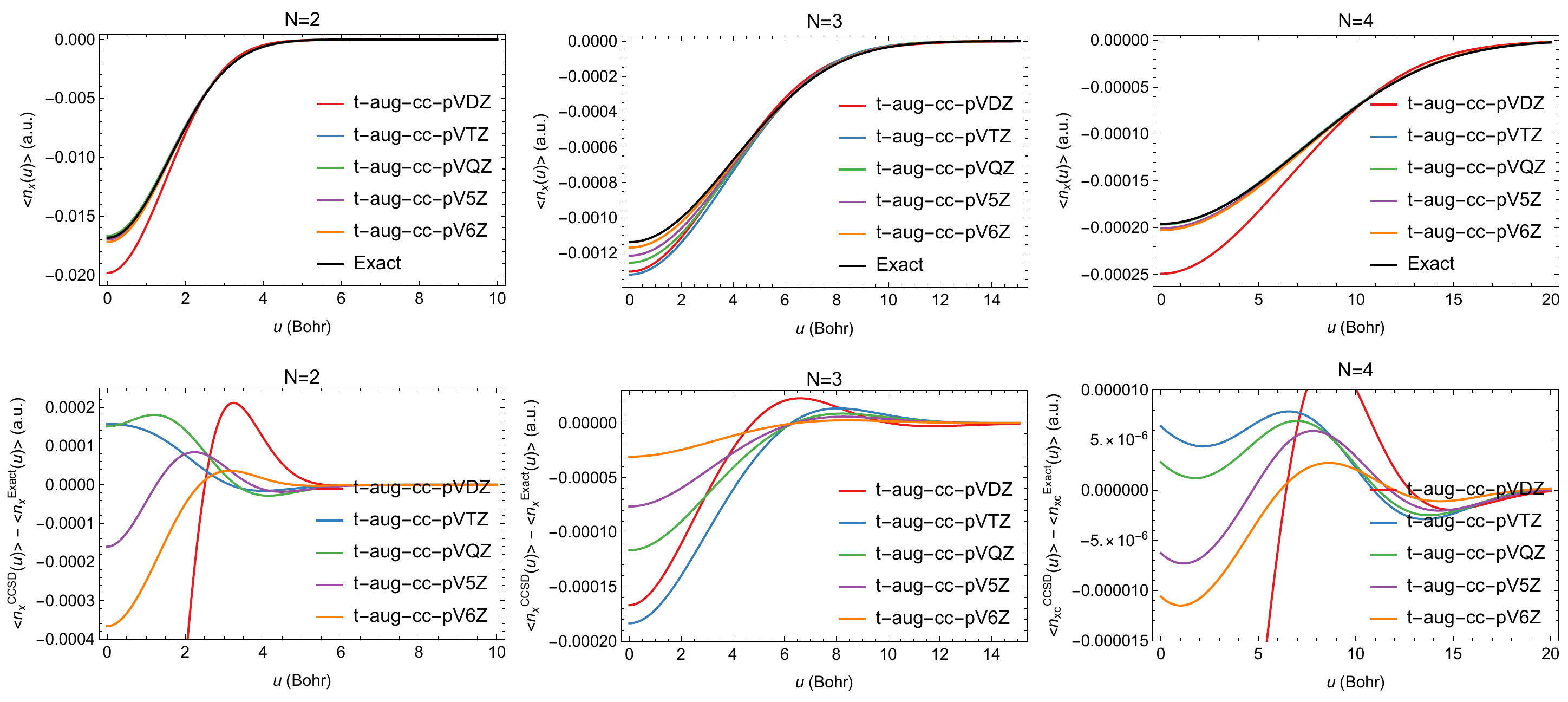}
\caption{System- and spherically-averaged exchange holes $\langle \nx(u) \rangle$ calculated with the Lieb functional at the CCSD level for the Hookium solutions with $N=2,\, 3,\, 4$ (upper panels) and the corresponding deviations with respect to exact results (lower panels). The orbital basis sets t-aug-cc-pV\textit{X}Z with $X=$D, T, Q, 5, 6 are employed for each $N$, with corresponding potential basis sets as described in Section~\ref{sec:computation}.}
\label{fig:Xhole_u}
\end{figure*}

\begin{table*}[!htp]\centering\renewcommand{\arraystretch}{1.24}
\setlength{\tabcolsep}{0pt}
\begin{threeparttable}
\caption{Percentage errors of the exchange energy $E_\xcx$ calculated by Lieb optimization at the CCSD level for the Hookium solutions with $N=2-11$. Orbital basis sets of $Y$-aug-cc-pV\textit{X}Z ($X=$ D, T, Q, 5, 6; $Y=$ d, t, q, p, s) and potential basis sets of ($Y-1$)-aug-cc-pV\textit{X}Z ($X=$ D, T, Q, 5, 6) are employed, respectively.}
\label{tab:PEs_Xene_ccsd}
\begin{tabular*}{.72\linewidth}{@{\extracolsep{\fill}}lzzzzzzzzzzz}
\toprule
\multicolumn{1}{c}{Orbital basis}&
\multicolumn{1}{c}{\,$N=2$\,} &
\multicolumn{1}{c}{\,$N=3$\,} &
  \multicolumn{1}{c}{\,$N=4$\,} &
  \multicolumn{1}{c}{\,$N=5$\,} &
  \multicolumn{1}{c}{$N=6$} &
  \multicolumn{1}{c}{$N=7$} &
  \multicolumn{1}{c}{$N=8$} &
  \multicolumn{1}{c}{$N=9$} &
  \multicolumn{1}{c}{$N=10$} &
  \multicolumn{1}{c}{$N=11$} \\ \midrule
d-aug-cc-pVDZ &
  \cellcolor[HTML]{C0D880}5.9\% &
  \cellcolor[HTML]{A4D07E}4.1\% &
  \cellcolor[HTML]{FFEB84}13.8\% &
  \cellcolor[HTML]{FFE082}70.4\% &
  \cellcolor[HTML]{FED280}142.2\% &
  \cellcolor[HTML]{FDC27D}227.0\% &
  \cellcolor[HTML]{FCAF79}324.3\% &
  \cellcolor[HTML]{FB9A75}434.0\% &
  \cellcolor[HTML]{FA8370}556.0\% &
  \cellcolor[HTML]{F8696B}690.1\% \\
d-aug-cc-pVTZ &
  \cellcolor[HTML]{6AC07B}0.5\% &
  \cellcolor[HTML]{BFD880}5.9\% &
  \cellcolor[HTML]{68BF7B}0.3\% &
  \cellcolor[HTML]{FFE583}43.7\% &
  \cellcolor[HTML]{FFDA81}103.0\% &
  \cellcolor[HTML]{FECC7E}173.7\% &
  \cellcolor[HTML]{FDBD7B}255.0\% &
  \cellcolor[HTML]{FCAB78}346.6\% &
  \cellcolor[HTML]{FB9874}448.6\% &
  \cellcolor[HTML]{FA8270}560.7\% \\
d-aug-cc-pVQZ &
  \cellcolor[HTML]{72C27B}1.0\% &
  \cellcolor[HTML]{A2D07E}4.0\% &
  \cellcolor[HTML]{63BE7B}0.0\% &
  \cellcolor[HTML]{FFE583}41.8\% &
  \cellcolor[HTML]{FFDA81}100.2\% &
  \cellcolor[HTML]{FECD7F}169.7\% &
  \cellcolor[HTML]{FDBE7C}249.8\% &
  \cellcolor[HTML]{FCAC78}340.1\% &
  \cellcolor[HTML]{FB9975}440.6\% &
  \cellcolor[HTML]{FA8471}551.1\% \\
d-aug-cc-pV5Z &
  \cellcolor[HTML]{6AC07B}0.4\% &
  \cellcolor[HTML]{8DCA7D}2.7\% &
  \cellcolor[HTML]{6DC17B}0.7\% &
  \cellcolor[HTML]{FFE583}42.5\% &
  \cellcolor[HTML]{FFDA81}101.0\% &
  \cellcolor[HTML]{FECD7F}170.8\% &
  \cellcolor[HTML]{FDBD7C}251.1\% &
  \cellcolor[HTML]{FCAC78}341.8\% &
  \cellcolor[HTML]{FB9975}442.6\% &
  \cellcolor[HTML]{FA8471}553.5\% \\
d-aug-cc-pV6Z &
  \cellcolor[HTML]{6AC07B}0.5\% &
  \cellcolor[HTML]{73C27B}1.0\% &
  \cellcolor[HTML]{83C77C}2.0\% &
  \cellcolor[HTML]{FFE583}44.1\% &
  \cellcolor[HTML]{FFDA81}103.1\% &
  \cellcolor[HTML]{FECC7E}173.6\% &
  \cellcolor[HTML]{FDBD7C}254.7\% &
  \cellcolor[HTML]{FCAB78}346.3\% &
  \cellcolor[HTML]{FB9874}448.1\% &
  \cellcolor[HTML]{FA8270}560.2\% \\ \hline
t-aug-cc-pVDZ &
  \cellcolor[HTML]{BED880}5.8\% &
  \cellcolor[HTML]{92CB7D}3.0\% &
  \cellcolor[HTML]{CADB80}6.5\% &
  \cellcolor[HTML]{B7D67F}5.4\% &
  \cellcolor[HTML]{FFEA84}19.8\% &
  \cellcolor[HTML]{FFE283}58.4\% &
  \cellcolor[HTML]{FFD981}104.3\% &
  \cellcolor[HTML]{FECF7F}156.6\% &
  \cellcolor[HTML]{FDC47D}214.8\% &
  \cellcolor[HTML]{FDB87B}279.1\% \\
t-aug-cc-pVTZ &
  \cellcolor[HTML]{6AC07B}0.4\% &
  \cellcolor[HTML]{B6D67F}5.3\% &
  \cellcolor[HTML]{68BF7B}0.3\% &
  \cellcolor[HTML]{B3D57F}5.1\% &
  \cellcolor[HTML]{A5D17E}4.2\% &
  \cellcolor[HTML]{FFE784}34.5\% &
  \cellcolor[HTML]{FFE082}72.4\% &
  \cellcolor[HTML]{FED781}116.1\% &
  \cellcolor[HTML]{FECE7F}165.0\% &
  \cellcolor[HTML]{FDC47D}218.9\% \\
t-aug-cc-pVQZ &
  \cellcolor[HTML]{72C27B}1.0\% &
  \cellcolor[HTML]{9BCE7E}3.6\% &
  \cellcolor[HTML]{63BE7B}0.0\% &
  \cellcolor[HTML]{B4D57F}5.2\% &
  \cellcolor[HTML]{C1D980}6.0\% &
  \cellcolor[HTML]{FFE683}37.1\% &
  \cellcolor[HTML]{FFDF82}75.8\% &
  \cellcolor[HTML]{FED680}120.4\% &
  \cellcolor[HTML]{FECD7F}170.2\% &
  \cellcolor[HTML]{FDC27D}225.3\% \\
t-aug-cc-pV5Z &
  \cellcolor[HTML]{69C07B}0.4\% &
  \cellcolor[HTML]{87C87D}2.3\% &
  \cellcolor[HTML]{68BF7B}0.4\% &
  \cellcolor[HTML]{B7D67F}5.3\% &
  \cellcolor[HTML]{FFEB84}9.9\% &
  \cellcolor[HTML]{FFE583}43.0\% &
  \cellcolor[HTML]{FFDD82}83.6\% &
  \cellcolor[HTML]{FED480}130.3\% &
  \cellcolor[HTML]{FECB7E}182.4\% &
  \cellcolor[HTML]{FDC07C}239.9\% \\
t-aug-cc-pV6Z &
  \cellcolor[HTML]{6AC07B}0.5\% &
  \cellcolor[HTML]{70C17B}0.8\% &
  \cellcolor[HTML]{6DC17B}0.7\% &
  \cellcolor[HTML]{A2D07E}4.0\% &
  \cellcolor[HTML]{FFEA84}17.9\% &
  \cellcolor[HTML]{FFE383}54.7\% &
  \cellcolor[HTML]{FFDA81}99.1\% &
  \cellcolor[HTML]{FED17F}149.8\% &
  \cellcolor[HTML]{FDC67D}206.4\% &
  \cellcolor[HTML]{FDBA7B}268.8\% \\ \hline
q-aug-cc-pVDZ &
  \cellcolor[HTML]{BED880}5.7\% &
  \cellcolor[HTML]{8FCA7D}2.8\% &
  \cellcolor[HTML]{BCD780}5.7\% &
  \cellcolor[HTML]{ABD27F}4.6\% &
  \cellcolor[HTML]{93CC7D}3.1\% &
  \cellcolor[HTML]{ECE582}8.7\% &
  \cellcolor[HTML]{99CD7E}3.5\% &
  \cellcolor[HTML]{FFE884}25.6\% &
  \cellcolor[HTML]{FFE383}52.4\% &
  \cellcolor[HTML]{FFDE82}82.7\% \\
q-aug-cc-pVTZ &
  \cellcolor[HTML]{69C07B}0.4\% &
  \cellcolor[HTML]{B5D57F}5.2\% &
  \cellcolor[HTML]{68BF7B}0.3\% &
  \cellcolor[HTML]{B4D57F}5.1\% &
  \cellcolor[HTML]{86C87D}2.3\% &
  \cellcolor[HTML]{BDD880}5.7\% &
  \cellcolor[HTML]{CEDC81}6.8\% &
  \cellcolor[HTML]{E8E482}8.4\% &
  \cellcolor[HTML]{FFE884}30.0\% &
  \cellcolor[HTML]{FFE383}55.1\% \\
q-aug-cc-pVQZ &
  \cellcolor[HTML]{72C27B}1.0\% &
  \cellcolor[HTML]{9ACE7E}3.5\% &
  \cellcolor[HTML]{63BE7B}0.0\% &
  \cellcolor[HTML]{B3D57F}5.1\% &
  \cellcolor[HTML]{8AC97D}2.5\% &
  \cellcolor[HTML]{D6DF81}7.3\% &
  \cellcolor[HTML]{A0CF7E}3.9\% &
  \cellcolor[HTML]{FFEB84}13.6\% &
  \cellcolor[HTML]{FFE684}36.7\% &
  \cellcolor[HTML]{FFE183}63.4\% \\
q-aug-cc-pV5Z &
  \cellcolor[HTML]{69C07B}0.4\% &
  \cellcolor[HTML]{86C87D}2.3\% &
  \cellcolor[HTML]{67BF7B}0.3\% &
  \cellcolor[HTML]{B0D47F}4.9\% &
  \cellcolor[HTML]{82C67C}2.0\% &
  \cellcolor[HTML]{DFE282}7.9\% &
  \cellcolor[HTML]{7EC57C}1.7\% &
  \cellcolor[HTML]{FFE984}22.5\% &
  \cellcolor[HTML]{FFE483}48.2\% &
  \cellcolor[HTML]{FFDF82}77.4\% \\
q-aug-cc-pV6Z &
  \cellcolor[HTML]{6AC07B}0.5\% &
  \cellcolor[HTML]{6FC17B}0.8\% &
  \cellcolor[HTML]{69BF7B}0.4\% &
  \cellcolor[HTML]{98CD7E}3.4\% &
  \cellcolor[HTML]{8FCA7D}2.8\% &
  \cellcolor[HTML]{99CD7E}3.4\% &
  \cellcolor[HTML]{FFEB84}15.0\% &
  \cellcolor[HTML]{FFE683}41.2\% &
  \cellcolor[HTML]{FFE082}71.8\% &
  \cellcolor[HTML]{FFD981}106.1\% \\ \hline
p-aug-cc-pVDZ &
  \cellcolor[HTML]{BED880}5.7\% &
  \cellcolor[HTML]{8ECA7D}2.8\% &
  \cellcolor[HTML]{BAD780}5.5\% &
  \cellcolor[HTML]{A9D27F}4.5\% &
  \cellcolor[HTML]{7FC67C}1.8\% &
  \cellcolor[HTML]{E2E282}8.0\% &
  \cellcolor[HTML]{74C37C}1.1\% &
  \cellcolor[HTML]{B3D57F}5.0\% &
  \cellcolor[HTML]{FFEB84}11.5\% &
  \cellcolor[HTML]{B5D57F}5.2\% \\
p-aug-cc-pVTZ &
  \cellcolor[HTML]{69C07B}0.4\% &
  \cellcolor[HTML]{B4D57F}5.2\% &
  \cellcolor[HTML]{68BF7B}0.3\% &
  \cellcolor[HTML]{B4D57F}5.1\% &
  \cellcolor[HTML]{81C67C}1.9\% &
  \cellcolor[HTML]{C5DA80}6.2\% &
  \cellcolor[HTML]{C3D980}6.1\% &
  \cellcolor[HTML]{7AC47C}1.5\% &
  \cellcolor[HTML]{EAE482}8.5\% &
  \cellcolor[HTML]{FFEB84}11.7\% \\
p-aug-cc-pVQZ &
  \cellcolor[HTML]{72C27B}1.0\% &
  \cellcolor[HTML]{9ACD7E}3.5\% &
  \cellcolor[HTML]{63BE7B}0.0\% &
  \cellcolor[HTML]{B2D57F}5.0\% &
  \cellcolor[HTML]{81C67C}1.9\% &
  \cellcolor[HTML]{D7DF81}7.3\% &
  \cellcolor[HTML]{9DCE7E}3.7\% &
  \cellcolor[HTML]{75C37C}1.2\% &
  \cellcolor[HTML]{FFEB84}10.8\% &
  \cellcolor[HTML]{EDE582}8.7\% \\
p-aug-cc-pV5Z &
  \cellcolor[HTML]{69C07B}0.4\% &
  \cellcolor[HTML]{86C87D}2.3\% &
  \cellcolor[HTML]{67BF7B}0.3\% &
  \cellcolor[HTML]{AFD47F}4.8\% &
  \cellcolor[HTML]{71C27B}0.9\% &
  \cellcolor[HTML]{D7DF81}7.3\% &
  \cellcolor[HTML]{6EC17B}0.7\% &
  \cellcolor[HTML]{CFDD81}6.8\% &
  \cellcolor[HTML]{FFEB84}9.9\% &
  \cellcolor[HTML]{7CC57C}1.6\% \\
p-aug-cc-pV6Z &
  \cellcolor[HTML]{6AC07B}0.5\% &
  \cellcolor[HTML]{6EC17B}0.8\% &
  \cellcolor[HTML]{68BF7B}0.3\% &
  \cellcolor[HTML]{96CC7D}3.2\% &
  \cellcolor[HTML]{9CCE7E}3.6\% &
  \cellcolor[HTML]{99CD7E}3.4\% &
  \cellcolor[HTML]{B2D47F}5.0\% &
  \cellcolor[HTML]{E3E382}8.1\% &
  \cellcolor[HTML]{7EC67C}1.8\% &
  \cellcolor[HTML]{FFEA84}18.1\% \\ \hline
s-aug-cc-pVDZ &
  \cellcolor[HTML]{BED880}5.7\% &
  \cellcolor[HTML]{8ECA7D}2.8\% &
  \cellcolor[HTML]{BAD780}5.5\% &
  \cellcolor[HTML]{A9D27F}4.4\% &
  \cellcolor[HTML]{7CC57C}1.6\% &
  \cellcolor[HTML]{E0E282}7.9\% &
  \cellcolor[HTML]{6EC17B}0.7\% &
  \cellcolor[HTML]{C1D980}6.0\% &
  \cellcolor[HTML]{FFEB84}10.9\% &
  \cellcolor[HTML]{B3D57F}5.1\% \\
s-aug-cc-pVTZ &
  \cellcolor[HTML]{69C07B}0.4\% &
  \cellcolor[HTML]{B4D57F}5.2\% &
  \cellcolor[HTML]{68BF7B}0.3\% &
  \cellcolor[HTML]{B4D57F}5.1\% &
  \cellcolor[HTML]{80C67C}1.9\% &
  \cellcolor[HTML]{C6DA80}6.3\% &
  \cellcolor[HTML]{C2D980}6.0\% &
  \cellcolor[HTML]{6BC07B}0.6\% &
  \cellcolor[HTML]{F2E783}9.0\% &
  \cellcolor[HTML]{FFEB84}11.0\% \\
s-aug-cc-pVQZ &
  \cellcolor[HTML]{72C27B}1.0\% &
  \cellcolor[HTML]{9ACD7E}3.5\% &
  \cellcolor[HTML]{63BE7B}0.0\% &
  \cellcolor[HTML]{B2D57F}5.0\% &
  \cellcolor[HTML]{80C67C}1.8\% &
  \cellcolor[HTML]{D7DF81}7.3\% &
  \cellcolor[HTML]{9DCE7E}3.7\% &
  \cellcolor[HTML]{89C97D}2.4\% &
  \cellcolor[HTML]{FFEB84}10.6\% &
  \cellcolor[HTML]{E2E282}8.0\% \\
s-aug-cc-pV5Z &
  \cellcolor[HTML]{69C07B}0.4\% &
  \cellcolor[HTML]{86C87D}2.2\% &
  \cellcolor[HTML]{67BF7B}0.3\% &
  \cellcolor[HTML]{AFD37F}4.8\% &
  \cellcolor[HTML]{6EC17B}0.7\% &
  \cellcolor[HTML]{D5DF81}7.2\% &
  \cellcolor[HTML]{76C37C}1.2\% &
  \cellcolor[HTML]{D9E081}7.5\% &
  \cellcolor[HTML]{F3E783}9.1\% &
  \cellcolor[HTML]{97CD7E}3.3\% \\
s-aug-cc-pV6Z &
  \cellcolor[HTML]{6AC07B}0.5\% &
  \cellcolor[HTML]{6EC17B}0.8\% &
  \cellcolor[HTML]{68BF7B}0.3\% &
  \cellcolor[HTML]{95CC7D}3.2\% &
  \cellcolor[HTML]{9ECF7E}3.7\% &
  \cellcolor[HTML]{9ACD7E}3.5\% &
  \cellcolor[HTML]{BFD880}5.8\% &
  \cellcolor[HTML]{D8DF81}7.4\% &
  \cellcolor[HTML]{9ACD7E}3.5\% &
  \cellcolor[HTML]{FAE983}9.5\% \\
\midrule
\bottomrule
\end{tabular*}
%\begin{tablenotes}
%\item[a]\label{1} The Lieb optimization process could not get converged.
%\end{tablenotes}
\end{threeparttable}
\end{table*}

For closed-shell two-electron systems, the exchange energy is related to the Hartree energy as $E_\text{x}\ndep=-\frac{1}{2}E_\text{H}\ndep$ and this dominates the XC energy. In addition, the exchange hole is related to the electron density for the closed-shell two-electron system as $n_\text{x}(\mathbf{r},\mathbf{r}') = -n(\mathbf{r}')$. As a result, the convergence of the exchange hole with respect to basis set size is the same as that observed for the electron density. This can be seen in the upper panels of Figure~\ref{fig:Xhole_u}, Figure~\ref{fig:ccsd_density} and Table~\ref{tab:PEs_density_CCSD}. Furthermore, different convergence patterns are observed in Figure~\ref{fig:Xhole_u} for solutions of even and odd values of $N$. Plots of the deviation of the finite-basis exchange holes from the exact solutions in Figure~\ref{fig:Xhole_u} indicate that basis-set convergence is generally reached with the t-aug-cc-pV6Z basis set for solutions of $N=2,\,3,\,4$.

Table~\ref{tab:PEs_Xene_ccsd} presents the PEs of exchange energy $\Ex$ obtained using Lieb optimization at the CCSD level for solutions to the Hookium atom of different order $N$ with increasing basis set size. An initial observation that can be made is that different error characteristics are again exhibited for solutions with odd and even values $N$ respectively. Specifically, for $N=2,\,4$ the error is relatively small for all basis-sets with $X>2$ while for solutions with $N=3$ and $N=5$, the PEs are generally larger in magnitude by comparison. Secondly, for a solution with any given $N$, once the number of diffuse basis functions is sufficient, increasing the cardinal number of the basis set will not increase the accuracy of the energy. For solutions with $N=2,\, 3,\, 4,\, 5$, the accuracy does not significantly improve beyond $Y=$ d, t, q, q respectively. For $N\geq6$, the improvements in accuracy with increasing cardinal number ceases at $Y=$p.

\subsubsection{The correlation hole and the description of the cusp}\label{subsubsec:corr_hole}
\begin{figure*}[!htp]\centering 
\includegraphics[width=\textwidth]{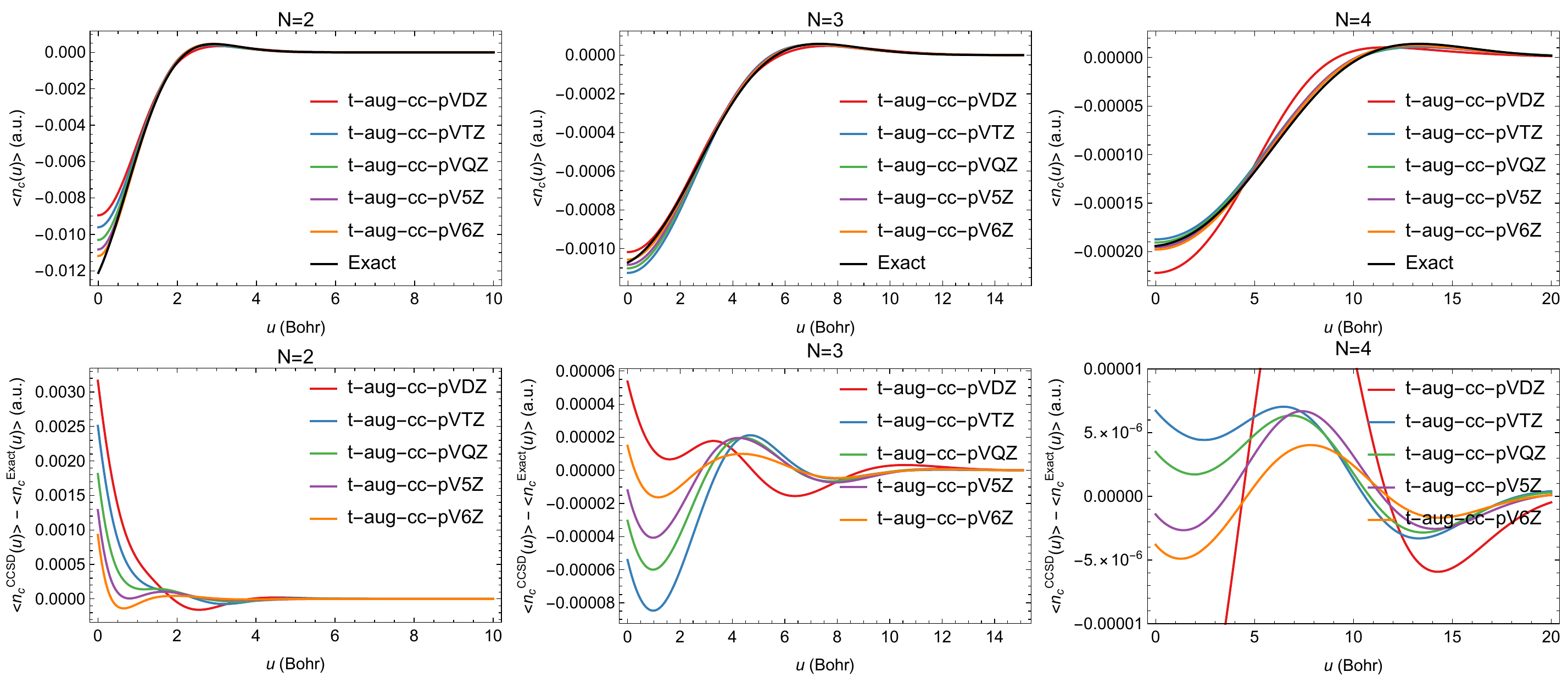}
\caption{System- and spherically-averaged correlation hole densities $\langle \nc^{\lambda=1}(u)\rangle$ calculated by Lieb optimization at the CCSD level (upper panels) for Hookium solutions with $N=2,\, 3,\, 4$ and corresponding errors with respect to exact results (lower panels).}
\label{fig:Chole_u}
\end{figure*}

%AMT: HERE CONTIUE THIS EVENING 29/3!
We now consider the system- and spherically-averaged correlation holes for Hookium atoms with $N=2,\,3,\,4$ in Figure~\ref{fig:Chole_u} and compare them with those of the corresponding exact solutions. 
%Subsequently, the electron-electron cusp condition associated with the finite-basis set treatment of the problem in CCSD is analyzed by quantifying the cusp-effect-driven errors in the correlation energy, which is illustrated in Table~\ref{tab:PEs_density_CCSD} and Figure~\ref{fig:Cene_u}. Finally, we tabulate the PEs of the correlation energy $E^{\lambda=1}_\xcc$ for different Hookium solutions with order $N$ for increasing basis-set sizes in Table\ref{tab:PEs_Cene_ccsd}.
Figure~\ref{fig:Chole_u} illustrates that, as the order of the Hookium solution $N$ increases from 2 to 4, the exact correlation holes become increasingly shallow. This trend is consistent with the behavior observed in the electron densities plotted in Figure~\ref{fig:ccsd_density} and the exchange holes in Figure~\ref{fig:Xhole_u}. In addition, the cusp at $u=0$ becomes shallower as the order of the solution $N$ increases, indicating that the cusp effect is less significant for more diffuse electron densities.

Figure~\ref{fig:Chole_u} displays the effect of basis set size on the correlation holes obtained via Lieb optimization at the CCSD level. With the exception of the t-aug-cc-pV6Z basis set for the most diffuse solution with $N=4$, enlarging the basis set results in an overall improved representation of the correlation holes with respect to those of the analytical solutions. In the lower panels of Figure~\ref{fig:Chole_u}, the error in the system- and spherically-averaged correlation holes are plotted radially from the atomic nuclei. Compared with higher-order solutions of larger $N$, the maximum error for $N=2$ arises at $u=0$, indicating that the cusp condition is more significant the more localized the electrons are, consistent with Eq.~(\ref{eq:cuspcond_Chole}). For the Hookium solution with $N=4$, the error is more uniformly distributed radially than for solutions with $N=2$ and $N=3$. 

\begin{table}[!ht]\centering\renewcommand{\arraystretch}{1.24}
\setlength{\tabcolsep}{0pt}
\begin{threeparttable}
\caption{The trends of the cusp-effect driven errors $\dWcPE$ defined by Eq.~\eqref{eq:cusp_err_def} in CCSD calculations for Hookium solutions with $N=2,\, 3,\, 4$, as the basis set size increases.}
\label{tab:cusp_errs}
\begin{tabular*}{\linewidth}{@{\extracolsep{\fill}}lzzz}
\toprule
\multicolumn{1}{c}{Orbital basis}
&\multicolumn{1}{c}{$N=2$}
&\multicolumn{1}{c}{$N=3$}
&\multicolumn{1}{c}{$N=4$} \\
\midrule
t-aug-cc-pVDZ  &12.67\%  & 2.65\%  &-0.08\% \\
t-aug-cc-pVTZ  & 9.49\%  &-0.76\%  & 0.23\% \\
t-aug-cc-pVQZ  & 4.99\%  &-0.53\%  & 0.08\% \\
t-aug-cc-pV5Z  & 0.86\%  &-0.35\%  &-0.05\% \\
t-aug-cc-pV6Z  &-0.55\%  &-0.18\%  &-0.09\% \\
\midrule
\bottomrule
\end{tabular*}
\end{threeparttable}
\end{table}
\begin{figure*}\centering
\includegraphics[width=\textwidth]{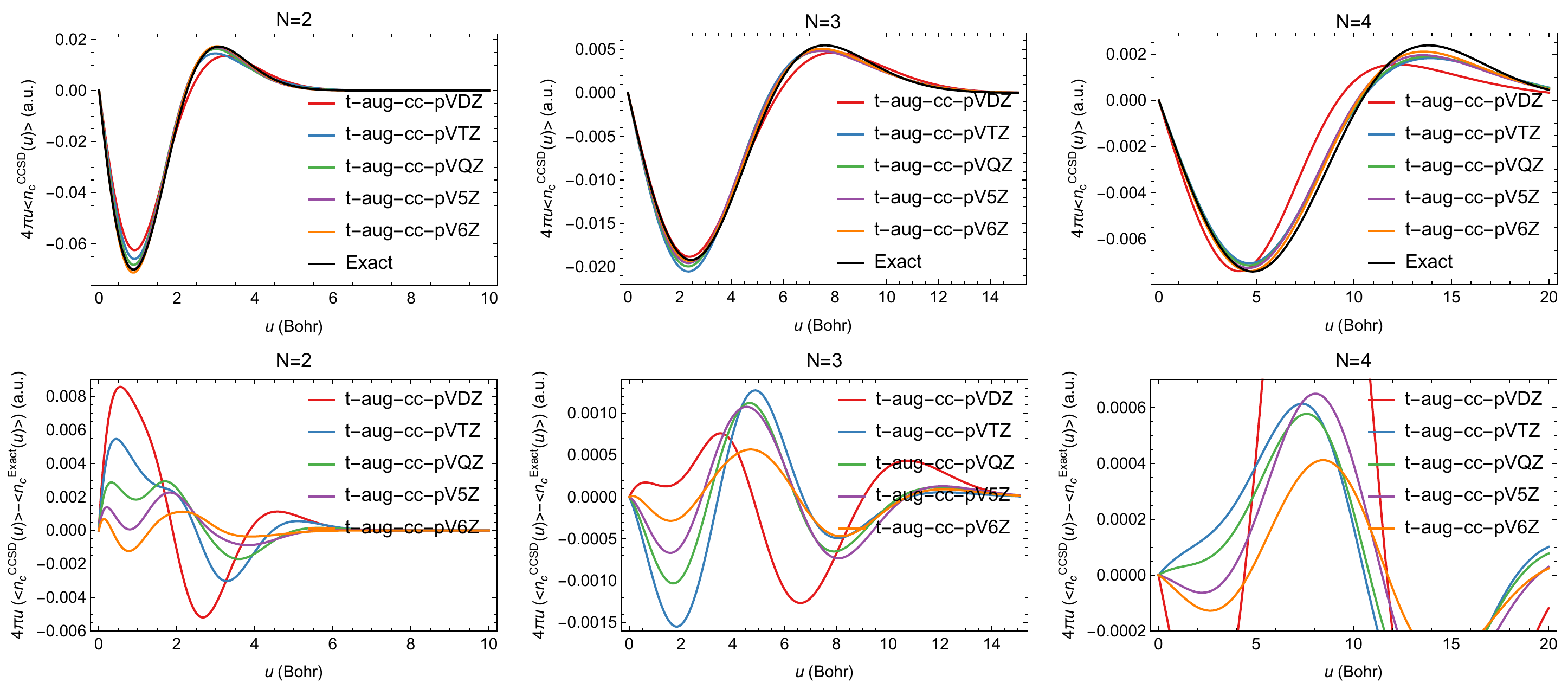}
\caption{%CCSD calculated correlation energy densities $\varepsilon^{\lambda=1}_\xcc(u)=4\pi u \langle n_c^{\lambda=1}(u)\rangle$ at $\lambda=1$ (upper panels) and their differences with respect to those of the analytical solutions for Hookium solutions of $N=2,\, 3,\, 4$. Orbital basis sets of t-aug-cc-pV\textit{X}Z ($X=$D, T, Q, 5, 6) are employed.}
CCSD evaulated $\varepsilon^{\lambda=1}_\xcc(u)=4\pi u \langle n_c^{\lambda=1}(u)\rangle$ (upper panels) and their differences with respect to exact results for Hookium solutions of $N=2,\, 3,\, 4$. Orbital basis sets of t-aug-cc-pV\textit{X}Z ($X=$D, T, Q, 5, 6) are employed.}
\label{fig:Cene_u}
\end{figure*}

To quantitatively estimate the effect of the electron-electron cusp, we define a cusp-effect driven error $\dWcPE$ in the correlation energy as
\begin{equation}
\dWcPE= \frac{ \displaystyle \int^{u_\xcc}_0\du\, 4\pi u \left\lbrack \langle \nc^\ccsd(u) \rangle -\langle \nc^\exact(u) \rangle \right\rbrack }{ \displaystyle \int^\infty_0\du\, 4\pi u \langle \nc^\exact(u) \rangle } \times 100\%
\label{eq:cusp_err_def}
\end{equation}
where a characteristic distance $u_\xcc$ defining the electron-electron cusp region is determined by the solution to
\begin{equation}
\left.\frac{\txtd}{\txtd u} \left\lbrack \langle \nc^\ccsd(u)\rangle -\langle \nc^\exact(u) \rangle \right\rbrack \right\vert_{u_\xcc}=0.
\end{equation}

Table~\ref{tab:cusp_errs} presents the cusp-effect driven errors $\dWcPE$ computed for Hookium solutions of $N=2,\,3,\,4$ with increasing basis set sizes. The trends of PEs with respect to $N$ are in agreement with the observations in Figure~\ref{fig:Chole_u}; solutions with higher $N$ values exhibit smaller errors resulting from the cusp effect. This is also consistent with the cusp condition for the correlation hole expressed in Eq.~\eqref{eq:cuspcond_Chole}. Since solutions of increasing $N$ have an increasingly diffuse electron density, as shown in Figure~\ref{fig:ccsd_density}, the integral $\int\dr\,n^2(\boldr)/(2\Ne)$ decreases, and $\langle \nc(0) \rangle$ becomes smaller~\cite{perdew1992pair}. This, in turn, leads to a reduction in $\langle \nc'(0)\rangle$, resulting in a flatter $\langle \nc(u) \rangle$ approaching $u=0$ for solutions of larger $N$, as demonstrated in Figure~\ref{fig:Chole_u}. It follows therefore that the cusp-driven error becomes much less significant when $N$ is large or the electron density is diffuse.

Table~\ref{tab:cusp_errs} also shows that, as the cardinal number \textit{X} of the basis set is increased from $3$ to $6$, the cusp errors become consistently smaller for solutions with $N=2$, $3$. However, the situation for the $N=4$ solution is different due to the cusp error being less significant given its diffuse electron density. With the largest t-aug-cc-pV6Z set considered in this comparison,  the cusp-driven errors $\dWcPE$ are $-0.55\%$, $-0.18\%$, $-0.09\%$ for $N=2$, $3$, $4$, respectively.

Figure~\ref{fig:Cene_u} plots the spherically-averaged correlation energy density with $\varepsilon^{\lambda=1}_\xcc(u)=4\pi u \langle n_c^{\lambda=1}(u)\rangle$. It shows that the cusp-effect driven error that arises at short interelectronic separations is significantly attenuated at larger values of the interelectronic distance $u$. Conversely, the errors in the correlation hole become more pronounced at larger values of $u$ in $\epsilon^{\lambda=1}_\xcc(u)$. Overall, calculations with the basis set with $X=6$ yield a markedly improved description of $\varepsilon^{\lambda=1}_\xcc(u)$ compared to the results obtained with smaller basis sizes for the less diffuse $N=2,\,3$ Hookium solutions. These observations validate the application of Lieb optimization at the CCSD level with the appropriate Gaussian basis sets to calculate XC holes accurately.

\begin{table*}[!htp]\centering\renewcommand{\arraystretch}{1.24}
\setlength{\tabcolsep}{0pt}
\begin{threeparttable}
\caption{Absolute percentage errors of the correlation energy $E^{\lambda=1}_\xcc$ % at $\lambda=1$ 
using Lieb optimization at the CCSD level for Hookium solutions with $N=2-11$. Orbital basis sets of $Y$-aug-cc-pV\textit{X}Z ($X=$D, T, Q, 5, 6; $Y=$d, t, q, p, s) and potential basis sets of ($Y-1$)-aug-cc-pV\textit{X}Z ($X=$D, T, Q, 5, 6) are employed.}
\label{tab:PEs_Cene_ccsd}
\begin{tabular*}{.72\linewidth}{@{\extracolsep{\fill}}lzzzzzzzzzzz}
\toprule
\multicolumn{1}{c}{Orbital basis}&
  \multicolumn{1}{c}{$\,N=2\,$} &
  \multicolumn{1}{c}{$\,N=3\,$} &
  \multicolumn{1}{c}{$N=4$} &
  \multicolumn{1}{c}{$N=5$} &
  \multicolumn{1}{c}{$N=6$} &
  \multicolumn{1}{c}{$N=7$} &
  \multicolumn{1}{c}{$N=8$} &
  \multicolumn{1}{c}{$N=9$} &
  \multicolumn{1}{c}{$N=10$} &
  \multicolumn{1}{c}{$N=11$} \\ \midrule
d-aug-cc-pVDZ &
  \cellcolor[HTML]{85C77C}5.1\% &
  \cellcolor[HTML]{7EC57C}4.2\% &
  \cellcolor[HTML]{FEC97E}46.1\% &
  \cellcolor[HTML]{FED380}39.0\% &
  \cellcolor[HTML]{FFEB84}22.0\% &
  \cellcolor[HTML]{69BF7B}1.2\% &
  \cellcolor[HTML]{FFEA84}23.1\% &
  \cellcolor[HTML]{FDC37D}50.3\% &
  \cellcolor[HTML]{FB9874}80.3\% &
  \cellcolor[HTML]{F8696B}113.0\% \\
d-aug-cc-pVTZ &
  \cellcolor[HTML]{92CB7D}7.0\% &
  \cellcolor[HTML]{8FCA7D}6.6\% &
  \cellcolor[HTML]{FCA477}71.8\% &
  \cellcolor[HTML]{FB9A75}79.0\% &
  \cellcolor[HTML]{FB9F76}75.3\% &
  \cellcolor[HTML]{FCA877}69.4\% &
  \cellcolor[HTML]{FCB279}62.2\% &
  \cellcolor[HTML]{FDBE7C}54.0\% &
  \cellcolor[HTML]{FECB7E}44.9\% &
  \cellcolor[HTML]{FFD981}34.9\% \\
d-aug-cc-pVQZ &
  \cellcolor[HTML]{90CB7D}6.6\% &
  \cellcolor[HTML]{7DC57C}4.0\% &
  \cellcolor[HTML]{FCAF79}64.2\% &
  \cellcolor[HTML]{FCA377}72.5\% &
  \cellcolor[HTML]{FCAA78}67.6\% &
  \cellcolor[HTML]{FDB67A}59.8\% &
  \cellcolor[HTML]{FDC37D}50.2\% &
  \cellcolor[HTML]{FED380}39.4\% &
  \cellcolor[HTML]{FFE483}27.4\% &
  \cellcolor[HTML]{C7DA80}14.3\% \\
d-aug-cc-pV5Z &
  \cellcolor[HTML]{80C67C}4.4\% &
  \cellcolor[HTML]{68BF7B}1.1\% &
  \cellcolor[HTML]{FDBD7C}54.4\% &
  \cellcolor[HTML]{FCB37A}61.8\% &
  \cellcolor[HTML]{FDBE7C}54.2\% &
  \cellcolor[HTML]{FECE7F}43.0\% &
  \cellcolor[HTML]{FFE182}29.4\% &
  \cellcolor[HTML]{C5DA80}14.0\% &
  \cellcolor[HTML]{77C37C}3.1\% &
  \cellcolor[HTML]{FCEA83}21.7\% \\
d-aug-cc-pV6Z &
  \cellcolor[HTML]{68BF7B}1.1\% &
  \cellcolor[HTML]{65BE7B}0.7\% &
  \cellcolor[HTML]{FDC37D}50.7\% &
  \cellcolor[HTML]{FDB87B}58.1\% &
  \cellcolor[HTML]{FDC47D}49.8\% &
  \cellcolor[HTML]{FED580}37.4\% &
  \cellcolor[HTML]{FFEB84}22.5\% &
  \cellcolor[HTML]{89C97D}5.7\% &
  \cellcolor[HTML]{BED880}13.0\% &
  \cellcolor[HTML]{FFDB81}33.5\% \\ \hline
t-aug-cc-pVDZ &
  \cellcolor[HTML]{85C77C}5.1\% &
  \cellcolor[HTML]{75C37C}2.8\% &
  \cellcolor[HTML]{A7D17E}9.8\% &
  \cellcolor[HTML]{FFE583}26.9\% &
  \cellcolor[HTML]{FFE683}25.8\% &
  \cellcolor[HTML]{B2D47F}11.4\% &
  \cellcolor[HTML]{9BCE7E}8.2\% &
  \cellcolor[HTML]{FFDF82}31.0\% &
  \cellcolor[HTML]{FDBA7B}56.6\% &
  \cellcolor[HTML]{FB9273}84.7\% \\
t-aug-cc-pVTZ &
  \cellcolor[HTML]{91CB7D}6.8\% &
  \cellcolor[HTML]{88C87D}5.5\% &
  \cellcolor[HTML]{6EC17B}1.9\% &
  \cellcolor[HTML]{FDC37D}50.2\% &
  \cellcolor[HTML]{FB9273}84.3\% &
  \cellcolor[HTML]{FA8E73}87.1\% &
  \cellcolor[HTML]{FB9073}86.0\% &
  \cellcolor[HTML]{FB9374}83.7\% &
  \cellcolor[HTML]{FB9774}80.9\% &
  \cellcolor[HTML]{FB9C75}77.7\% \\
t-aug-cc-pVQZ &
  \cellcolor[HTML]{8FCA7D}6.4\% &
  \cellcolor[HTML]{78C47C}3.3\% &
  \cellcolor[HTML]{74C37C}2.8\% &
  \cellcolor[HTML]{FDC57D}48.9\% &
  \cellcolor[HTML]{FB9874}80.6\% &
  \cellcolor[HTML]{FB9474}83.2\% &
  \cellcolor[HTML]{FB9774}81.5\% &
  \cellcolor[HTML]{FB9B75}78.5\% &
  \cellcolor[HTML]{FBA076}74.7\% &
  \cellcolor[HTML]{FCA677}70.4\% \\
t-aug-cc-pV5Z &
  \cellcolor[HTML]{7FC67C}4.3\% &
  \cellcolor[HTML]{67BF7B}0.9\% &
  \cellcolor[HTML]{7BC57C}3.8\% &
  \cellcolor[HTML]{FECC7F}43.8\% &
  \cellcolor[HTML]{FCA878}69.1\% &
  \cellcolor[HTML]{FCA777}69.7\% &
  \cellcolor[HTML]{FCAD79}65.4\% &
  \cellcolor[HTML]{FDB67A}59.2\% &
  \cellcolor[HTML]{FDC17C}51.8\% &
  \cellcolor[HTML]{FECD7F}43.4\% \\
t-aug-cc-pV6Z &
  \cellcolor[HTML]{68BF7B}1.1\% &
  \cellcolor[HTML]{65BE7B}0.7\% &
  \cellcolor[HTML]{6FC17B}2.0\% &
  \cellcolor[HTML]{FECB7E}44.7\% &
  \cellcolor[HTML]{FCB37A}61.5\% &
  \cellcolor[HTML]{FDB67A}59.4\% &
  \cellcolor[HTML]{FDC07C}52.4\% &
  \cellcolor[HTML]{FECD7F}43.4\% &
  \cellcolor[HTML]{FFDC82}32.8\% &
  \cellcolor[HTML]{F7E883}21.0\% \\ \hline
q-aug-cc-pVDZ &
  \cellcolor[HTML]{85C77C}5.1\% &
  \cellcolor[HTML]{73C27B}2.6\% &
  \cellcolor[HTML]{A6D17E}9.7\% &
  \cellcolor[HTML]{FEEA83}22.0\% &
  \cellcolor[HTML]{F2E783}20.3\% &
  \cellcolor[HTML]{FFE483}27.4\% &
  \cellcolor[HTML]{FDEA83}21.8\% &
  \cellcolor[HTML]{9ECF7E}8.6\% &
  \cellcolor[HTML]{98CD7E}7.8\% &
  \cellcolor[HTML]{FFE583}26.3\% \\
q-aug-cc-pVTZ &
  \cellcolor[HTML]{91CB7D}6.8\% &
  \cellcolor[HTML]{87C87D}5.4\% &
  \cellcolor[HTML]{6CC07B}1.6\% &
  \cellcolor[HTML]{FFEB84}22.2\% &
  \cellcolor[HTML]{69BF7B}1.2\% &
  \cellcolor[HTML]{FDC07C}52.3\% &
  \cellcolor[HTML]{FA8C72}88.8\% &
  \cellcolor[HTML]{FA8270}95.7\% &
  \cellcolor[HTML]{FA8270}95.5\% &
  \cellcolor[HTML]{FA8370}94.9\% \\
q-aug-cc-pVQZ &
  \cellcolor[HTML]{8ECA7D}6.4\% &
  \cellcolor[HTML]{77C37C}3.2\% &
  \cellcolor[HTML]{70C17B}2.2\% &
  \cellcolor[HTML]{EBE582}19.4\% &
  \cellcolor[HTML]{74C27B}2.7\% &
  \cellcolor[HTML]{FDB47A}60.9\% &
  \cellcolor[HTML]{FA8B72}89.4\% &
  \cellcolor[HTML]{FA8471}94.2\% &
  \cellcolor[HTML]{FA8471}94.1\% &
  \cellcolor[HTML]{FA8571}93.4\% \\
q-aug-cc-pV5Z &
  \cellcolor[HTML]{7FC67C}4.3\% &
  \cellcolor[HTML]{67BF7B}0.9\% &
  \cellcolor[HTML]{76C37C}3.0\% &
  \cellcolor[HTML]{E8E482}18.9\% &
  \cellcolor[HTML]{8DCA7D}6.3\% &
  \cellcolor[HTML]{FDB77A}59.1\% &
  \cellcolor[HTML]{FB9B75}78.0\% &
  \cellcolor[HTML]{FB9874}80.7\% &
  \cellcolor[HTML]{FB9975}79.8\% &
  \cellcolor[HTML]{FB9C75}77.5\% \\
q-aug-cc-pV6Z &
  \cellcolor[HTML]{68BF7B}1.1\% &
  \cellcolor[HTML]{65BE7B}0.7\% &
  \cellcolor[HTML]{70C17B}2.2\% &
  \cellcolor[HTML]{BFD880}13.2\% &
  \cellcolor[HTML]{F1E783}20.2\% &
  \cellcolor[HTML]{FDB87B}58.2\% &
  \cellcolor[HTML]{FCAD78}65.9\% &
  \cellcolor[HTML]{FCAE79}64.8\% &
  \cellcolor[HTML]{FDB47A}60.9\% &
  \cellcolor[HTML]{FDBC7B}55.4\% \\ \hline
p-aug-cc-pVDZ &
  \cellcolor[HTML]{85C77C}5.1\% &
  \cellcolor[HTML]{73C27B}2.6\% &
  \cellcolor[HTML]{A7D17E}9.8\% &
  \cellcolor[HTML]{FCEA83}21.7\% &
  \cellcolor[HTML]{F8E983}21.1\% &
  \cellcolor[HTML]{FFE683}25.8\% &
  \cellcolor[HTML]{FFE984}23.9\% &
  \cellcolor[HTML]{FED07F}41.3\% &
  \cellcolor[HTML]{FED680}37.1\% &
  \cellcolor[HTML]{FFE984}24.1\% \\
p-aug-cc-pVTZ &
  \cellcolor[HTML]{91CB7D}6.8\% &
  \cellcolor[HTML]{87C87D}5.3\% &
  \cellcolor[HTML]{6BC07B}1.5\% &
  \cellcolor[HTML]{F6E883}20.9\% &
  \cellcolor[HTML]{64BE7B}0.5\% &
  \cellcolor[HTML]{FFE082}30.0\% &
  \cellcolor[HTML]{C1D980}13.5\% &
  \cellcolor[HTML]{99CD7E}7.9\% &
  \cellcolor[HTML]{FDB47A}60.7\% &
  \cellcolor[HTML]{FA8771}92.3\% \\
p-aug-cc-pVQZ &
  \cellcolor[HTML]{8ECA7D}6.4\% &
  \cellcolor[HTML]{77C37C}3.2\% &
  \cellcolor[HTML]{70C17B}2.2\% &
  \cellcolor[HTML]{E1E282}18.0\% &
  \cellcolor[HTML]{6EC17B}2.0\% &
  \cellcolor[HTML]{FFE784}25.5\% &
  \cellcolor[HTML]{B9D67F}12.3\% &
  \cellcolor[HTML]{FFE182}29.7\% &
  \cellcolor[HTML]{FB9975}79.9\% &
  \cellcolor[HTML]{FA7D6F}99.5\% \\
p-aug-cc-pV5Z &
  \cellcolor[HTML]{7FC67C}4.3\% &
  \cellcolor[HTML]{66BF7B}0.9\% &
  \cellcolor[HTML]{75C37C}2.9\% &
  \cellcolor[HTML]{DDE182}17.3\% &
  \cellcolor[HTML]{84C77C}5.0\% &
  \cellcolor[HTML]{FFE884}24.7\% &
  \cellcolor[HTML]{ADD37F}10.7\% &
  \cellcolor[HTML]{FEC87E}47.0\% &
  \cellcolor[HTML]{FB9C75}77.9\% &
  \cellcolor[HTML]{FA8C72}89.1\% \\
p-aug-cc-pV6Z &
  \cellcolor[HTML]{68BF7B}1.1\% &
  \cellcolor[HTML]{66BE7B}0.7\% &
  \cellcolor[HTML]{71C27B}2.3\% &
  \cellcolor[HTML]{B5D57F}11.7\% &
  \cellcolor[HTML]{D9E081}16.8\% &
  \cellcolor[HTML]{B6D67F}11.9\% &
  \cellcolor[HTML]{FFE684}25.7\% &
  \cellcolor[HTML]{FDB87B}58.2\% &
  \cellcolor[HTML]{FCA978}68.9\% &
  \cellcolor[HTML]{FCA677}70.6\% \\ \hline
s-aug-cc-pVDZ &
  \cellcolor[HTML]{85C77C}5.1\% &
  \cellcolor[HTML]{73C27B}2.6\% &
  \cellcolor[HTML]{A7D17E}9.8\% &
  \cellcolor[HTML]{FBEA83}21.6\% &
  \cellcolor[HTML]{F9E983}21.3\% &
  \cellcolor[HTML]{FFE784}25.5\% &
  \cellcolor[HTML]{FFE884}24.7\% &
  \cellcolor[HTML]{FED07F}41.1\% &
  \cellcolor[HTML]{FED881}35.6\% &
  \cellcolor[HTML]{FFE884}24.6\% \\
s-aug-cc-pVTZ &
  \cellcolor[HTML]{91CB7D}6.8\% &
  \cellcolor[HTML]{87C87D}5.3\% &
  \cellcolor[HTML]{6BC07B}1.5\% &
  \cellcolor[HTML]{F5E883}20.7\% &
  \cellcolor[HTML]{63BE7B}0.3\% &
  \cellcolor[HTML]{FFE283}28.8\% &
  \cellcolor[HTML]{BED880}13.0\% &
  \cellcolor[HTML]{A5D17E}9.6\% &
  \cellcolor[HTML]{FED380}39.3\% &
  \cellcolor[HTML]{FFE383}28.2\% \\
s-aug-cc-pVQZ &
  \cellcolor[HTML]{8ECA7D}6.4\% &
  \cellcolor[HTML]{77C37C}3.2\% &
  \cellcolor[HTML]{70C17B}2.2\% &
  \cellcolor[HTML]{E0E282}17.8\% &
  \cellcolor[HTML]{6FC17B}2.1\% &
  \cellcolor[HTML]{FFE884}24.2\% &
  \cellcolor[HTML]{ABD27F}10.4\% &
  \cellcolor[HTML]{F8E983}21.2\% &
  \cellcolor[HTML]{FFE082}30.1\% &
  \cellcolor[HTML]{FFEA84}23.3\% \\
s-aug-cc-pV5Z &
  \cellcolor[HTML]{7FC67C}4.3\% &
  \cellcolor[HTML]{66BF7B}0.8\% &
  \cellcolor[HTML]{75C37C}2.9\% &
  \cellcolor[HTML]{DBE081}17.1\% &
  \cellcolor[HTML]{86C87D}5.2\% &
  \cellcolor[HTML]{FFEA84}22.8\% &
  \cellcolor[HTML]{93CC7D}7.1\% &
  \cellcolor[HTML]{FFDF82}30.5\% &
  \cellcolor[HTML]{FFE283}28.5\% &
  \cellcolor[HTML]{E8E482}18.9\% \\
s-aug-cc-pV6Z &
  \cellcolor[HTML]{68BF7B}1.1\% &
  \cellcolor[HTML]{66BE7B}0.7\% &
  \cellcolor[HTML]{71C27B}2.3\% &
  \cellcolor[HTML]{B3D57F}11.5\% &
  \cellcolor[HTML]{D6DF81}16.5\% &
  \cellcolor[HTML]{B0D47F}11.1\% &
  \cellcolor[HTML]{FFEA84}23.0\% &
  \cellcolor[HTML]{FFEA84}22.8\% &
  \cellcolor[HTML]{BED880}13.0\% &
  \cellcolor[HTML]{FED07F}41.4\% \\
\midrule
\bottomrule
\end{tabular*}
\end{threeparttable}
\end{table*}
Table~\ref{tab:PEs_Cene_ccsd} collects the PEs of the correlation energy $E^{\lambda=1}_\xcc$ calculated via Lieb optimization at the CCSD level, by subtracting the Lieb $\lambda=0$ energy from the CCSD energy. Similar trends are observed for the correlation energy with respect to Hookium atom solution $N$ and basis set size, compared with $E_\xcx$ shown in Table~\ref{tab:PEs_Xene_ccsd}. While the greatest accuracy is obtained for Hookium atom solutions with $N\leq 4$, the PEs for the correlation energy are usually $3$--$7$ times larger than those for $E_\xcx$ in the same calculation. For $N=2$, $3$, $4$, the $Y$-aug-cc-pV6Z basis sets ($Y=$t, q, p, s) consistently yield accurate correlation energies with PEs of $1$--$2\%$. However, as $N$ increases beyond 4, the improvement in accuracy achieved by increasing the basis set size (either via larger cardinal numbers \textit{X} or increased augmentation $Y$) is not as significant as was observed for $E_\xcx$. 

\subsubsection{The exchange and correlation hole}\label{subsubsec:xc_hole}
\begin{figure*}[!htp]\centering 
\includegraphics[width=\textwidth]{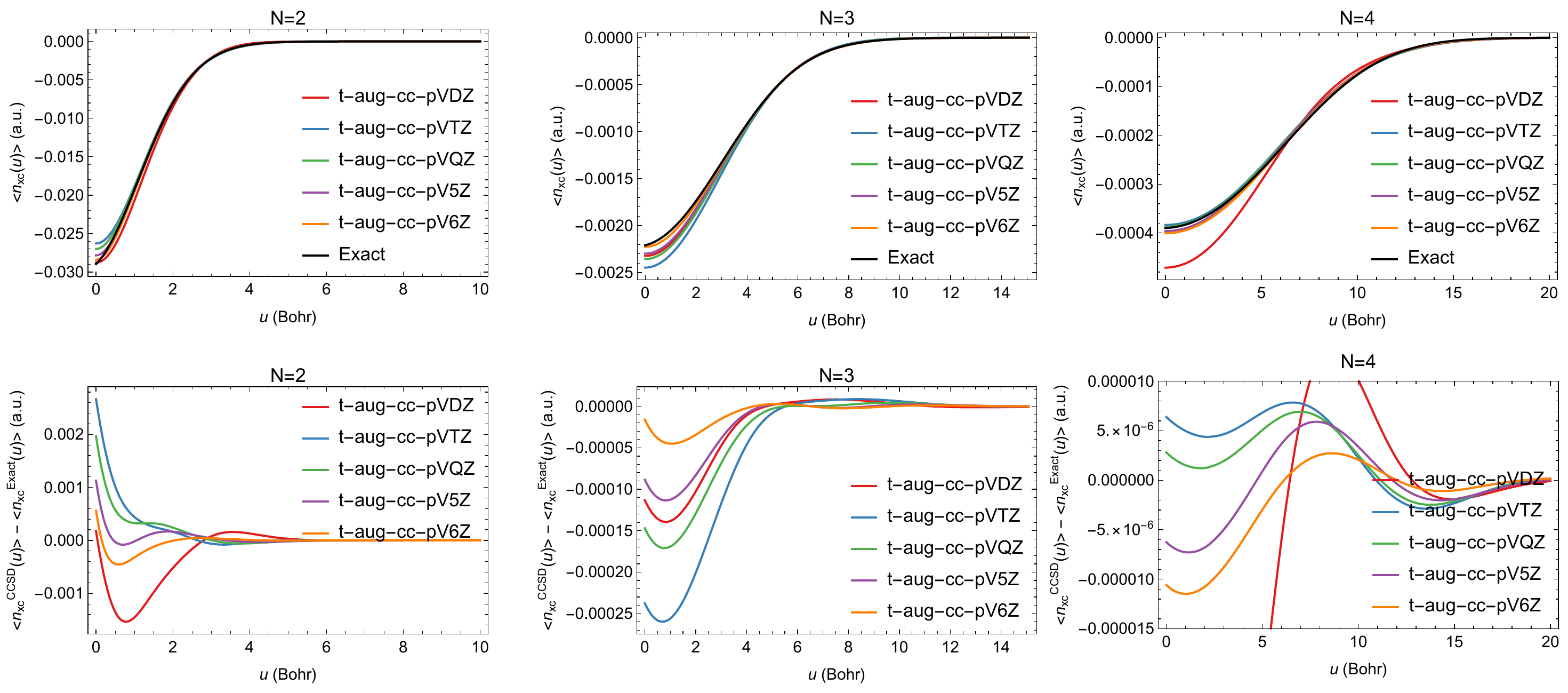}
\caption{The System- and spherically-averaged XC hole density $\langle \nxc^{\lambda=1}(u)\rangle$ calculated at the CCSD level (upper panels) for Hookium solutions with $N=2$, 3, 4 in the t-aug-cc-pV\textit{X}Z ($X=$D, T, Q, 5, 6) orbital basis sets, and the deviations with respect to those of the analytical solutions (lower panels).}
\label{fig:XChole_u}
\end{figure*}

Figure~\ref{fig:XChole_u} shows the XC holes $\nxc^{\lambda=1}(u)$ for Hookium atom solutions with $N=2$, $3$, $4$. Convergence of the XC holes with respect to the basis set is smoothly achieved for all three cases. In the case of $N=2$, $3$, the correlation component of the XC holes have significant errors due to the electron-electron cusp, and increasing the cardinal number 
 of the basis set leads to improved accuracy. It is worth noting that for the $N=2$ solution, the correlation hole around $u=0$ from Lieb optimization at the CCSD level are too shallow, which is compensated by the X hole being too deep, with the resulting error cancellation yielding a better accuracy for the XC hole than for either component individually. Overall, Lieb optimization at the CCSD level in the largest basis set t-aug-cc-pV6Z provides a satisfactory description of the XC hole for both the $N=2$ and $N=3$ Hookium atom solutions, for which the electron densities are relatively localized and not too diffuse.

\subsection{Coupling-constant averaged XC holes and hole models}\label{subsec:DFT_XCholes}
To assess the quality of the $\lambda$-averaged DFT XC holes, we first analyze how closely they align with the Lieb optimization results at the CCSD level calculated in the same basis set. This is important because CCSD-based Lieb optimizations have a much higher computational cost than DFT calculations and hence have a much greater limitation in terms of system and basis set size.

\begin{figure*}[!htp]\centering 
\includegraphics[width=\textwidth]{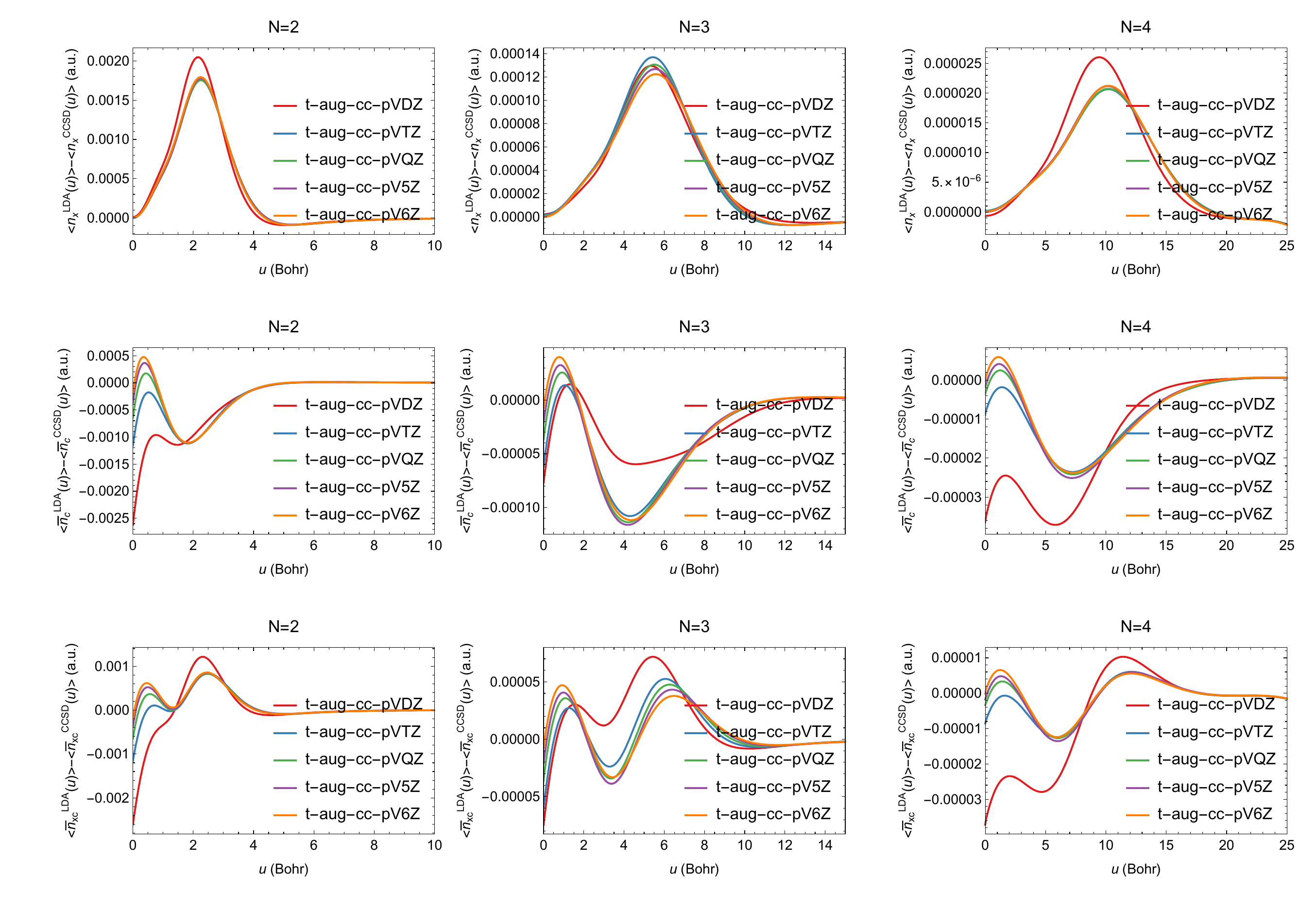}
\caption{Deviations of LDA hole densities $\langle \nx^\lda(u)\rangle,\, \langle \bar{n}^\lda_\xcc(u)\rangle$ and $\langle \bar{n}^\lda_\xc(u)\rangle$ from the corresponding Lieb optimization results at the CCSD level employing the same basis set for Hookium solutions with $N=2$, 3, 4.}
\label{fig:XCholeLDA_errfunc_basissize}
\end{figure*}

In Figure~\ref{fig:XCholeLDA_errfunc_basissize}, we have plotted the accuracy of LDA hole densities evaluated by $\langle n^\lda_\xcx(u) \rangle - \langle n^\ccsd_\xcx(u) \rangle$, $\langle \bar{n}^\lda_\xcc(u) \rangle -\langle \bar{n}^\ccsd_\xcc(u) \rangle$ and $\langle \bar{n}^\lda_\xc(u) \rangle -\langle \bar{n}^\ccsd_\xc(u) \rangle$. The cardinal number \textit{X} of the basis set was increased continuously from $2$ to $6$ to examine the convergence of this error with respect to basis set size.  
It is important to note that the errors corresponding to the $X=2$ basis set show a different pattern of errors to those of the larger basis sets, indicating this small basis is typically insufficient to accurately evaluate the $\lambda$-averaged XC holes. This is consistent with the previous discussion concerning the $\lambda=1$ case. 

Figure~\ref{fig:XCholeLDA_errfunc_basissize} shows that the largest change in $\langle n^\lda_\xcx(u) \rangle - \langle n^\ccsd_\xcx(u) \rangle$ with respect to the basis set size occurs at the value of $u$ with the largest error, particularly in the case of the $N=3$ Hookium atom solution. Nevertheless, overall the LDA hole model calculations of the exchange holes exhibit a rapid convergence in their errors with respect to basis set size, suggesting that the accuracy of the DFT exchange holes is relatively insensitive to basis set size.

However, for the $\lambda$-averaged correlation holes, the situation is somewhat different. Changes in the error of the LDA correlation holes with respect to increasing basis set size are most visible around the first peak in the radial plots of the correlation holes, extending to subsequent peaks further from the nucleus for the $N=3,\,4$ Hookium atom solutions. 
As the basis set grows, the position of the first peak tends to shift toward $u=0$. This could be attributed to the fact that the cusp-driven deviations become less significant with larger basis sets with CCSD-based calculations, whereas LDA is designed to satisfy the cusp condition and converges much more rapidly with basis set size.

% Increasing the basis set size by changing the cardinal number from $2$ to $3$ results in a significant improvement in the representation of the correlation hole for the Hookium $N=2$ solution. However, for accurate representation of the correlation hole at short range with $u<1$ bohr, a basis with $X \geq 4$ would appear to be required.

% Concerning the $\lambda$-averaged XC hole, the difference between results in different basis sets is largely dependent on the order of the Hookium solution $N$.  
% For the $N=2$ and $4$ solutions, these difference between basis sets of different size are mostly limited to the short range with $u\leq2$ bohr, while for the $N=3$ solution significant differences between basis sets are also distinguishable at a larger $u$ but with a gradually decreasing magnitude.

Although the exchange hole is the dominant component of the XC hole, the errors of the LDA exchange holes and correlation holes are comparable in size. In Figure~\ref{fig:XCholeLDA_errfunc_basissize}, the errors at short-range for the LDA XC holes are dominated by the correlation hole contribution, while errors at mid-range mainly arise from LDA exchange hole or both. Overall, accurate $\lambda$-averaged XC hole (or correlation hole) calculations require the use of basis-sets with $X \geq 4$. It is worth mentioning that similar trends are observed for the PBE hole model, the results of which are presented in the Supporting Information.

\begin{figure*}[!htp]\centering
\includegraphics[width=\textwidth]{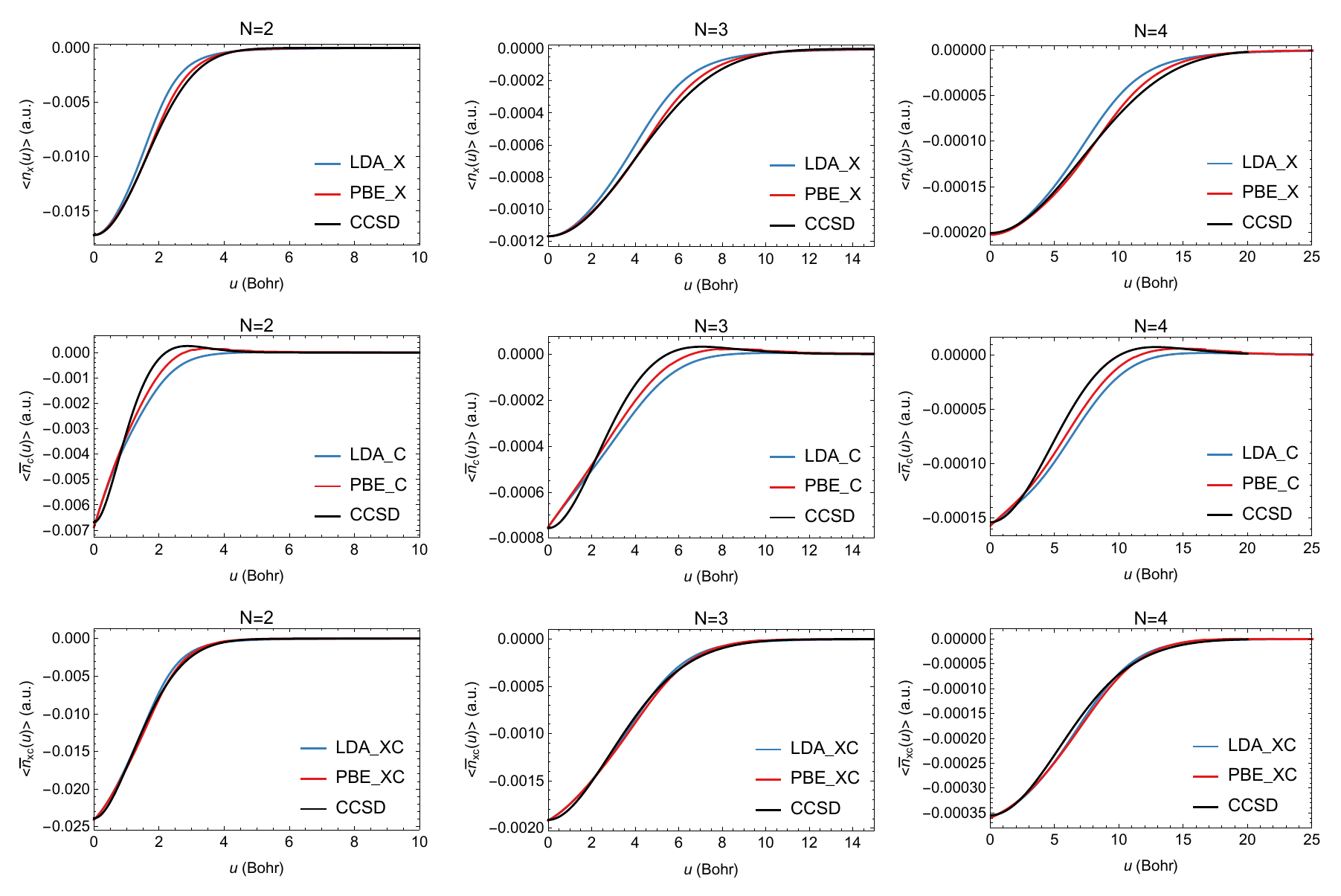}
\caption{System- and spherical-averaged $\langle \nx(u)\rangle$, and $\lambda$-averaged $\langle \bar{n}_{c}(u)\rangle$, $\langle \bar{n}_\xc(u)\rangle$ calculated by LDA and PBE hole models as compared with the Lieb optimization results at the CCSD level for Hookium solutions with $N=2,\,3,\,4$ in the t-aug-cc-pV6Z orbital basis set and d-aug-cc-pV6Z potential basis set.}
\label{fig:DFT_XChole_u}
\end{figure*}

\begin{table*}[!htp]\centering\renewcommand{\arraystretch}{1.24}
\setlength{\tabcolsep}{0pt}
\begin{threeparttable}
\caption{%Percentage errors of $E^\lda_\xcx$, $E^\lda_\xcc$ and $E^\lda_\xc$ calculated from the LDA hole model relative to the Lieb optimization results at the CCSD level with the same orbital basis set for Hookium atom solutions of $N=2,\,3,\,4$.
Percentage errors of energy components $E^\lda_\xcx$, $E^\lda_\xcc$ and $E^\lda_\xc$ calculated from LDA hole model relative to the CCSD+Lieb results under the same orbital basis set for Hookium solutions of $N=2,\,3,\,4$.}
\label{tab:PEs_Exc_LDA}
\begin{tabular*}{.8\linewidth}{@{\extracolsep{\fill}}lzzzzzzzzz}
\toprule
&\multicolumn{3}{c}{$\Ex^\lda$}&\multicolumn{3}{c}{$\Ec^\lda$}&\multicolumn{3}{c}{$\Exc^\lda$} \\
\cline{2-4}\cline{5-7}\cline{8-10}
$N$&\multicolumn{1}{c}{2} &\multicolumn{1}{c}{3} &\multicolumn{1}{c}{4}
&\multicolumn{1}{c}{2} &\multicolumn{1}{c}{3} &\multicolumn{1}{c}{4} &\multicolumn{1}{c}{2} &\multicolumn{1}{c}{3} &\multicolumn{1}{c}{4} \\
t-aug-cc-pVDZ
&  14.4\%  & 14.6\%  &  14.9\%
&-155.4\%  &-63.1\%  &-107.9\%
&   4.2\%  &  3.9\%  &  -1.8\%  \\
t-aug-cc-pVTZ
&  14.5\%  & 14.2\%  & 14.9\%
&-146.0\%  &-83.8\%  &-79.2\%
&   4.3\%  &  2.2\%  & -0.1\% \\
t-aug-cc-pVQZ
&  14.6\%  & 14.3\%  & 14.9\%
&-138.7\%  &-88.6\%  &-78.6\%
&   4.5\%  &  1.8\%  & -0.1\% \\
t-aug-cc-pV5Z
&  14.6\%  & 14.3\%  & 14.8\%
&-132.1\%  &-92.2\%  &-79.9\%
&   4.7\%  &  1.6\%  & -0.2\% \\
t-aug-cc-pV6Z
&  14.5\%  & 14.4\%  & 14.7\%
&-126.5\%  &-92.3\%  &-76.8\%
&   4.8\%  &  1.5\%  &  0.0\%  \\\midrule
\bottomrule
\end{tabular*}
\end{threeparttable}
\end{table*}

The XC holes obtained with different basis sets are used to calculate the corresponding LDA XC energies, and Table~\ref{tab:PEs_Exc_LDA} presents the PEs of the LDA XC energies relative to Lieb optimization values at the CCSD level for basis sets of increasing size. Table~\ref{tab:PEs_Exc_LDA} shows that the PE variations of the LDA exchange energy are relatively small, within $0.2\%$, while the convergence behavior of correlation energy for the Hookium solutions with $N=3,\,4$ exhibits no clear trend. However, for the $N=2$ solution, in which the cusp-effect driven error is the most significant in the CCSD calculations, the PE changes of the LDA correlation energy with increasing basis set size are slightly larger. The changes in the PE of $\Exc$ with increasing basis-set size are similar to that of the exchange energy, with only a change of $0.3\%$ for the $N=2$ solution from the smallest to the largest basis set; this is because $\Ex$ represents the vast majority of $\Exc$. 

In Figure~\ref{fig:DFT_XChole_u}, the exchange holes, $\lambda$-averaged correlation holes, and XC holes from the LDA hole model, PBE hole model, and Lieb optimizations at the CCSD level are presented with the t-aug-cc-pV6Z basis set. Figure~\ref{fig:DFT_XChole_u} shows that both the LDA and PBE hole models, in particular the LDA one, tend to localize the exchange hole, regardless of the order of the solution $N$. For the correlation holes, although both LDA and PBE hole models capture the cusp condition, they exhibit an almost linear behaviour before reaching their maximum value, resulting in an overly shallow correlation hole density in the small $u$ region but an overly deep correlation hole at the intermediate $u$ region. Figure~\ref{fig:DFT_XChole_u} also shows that, for both the exchange and correlation holes, the PBE hole model is superior to the corresponding LDA hole model. However, the LDA and PBE model XC holes appear much more similar due to error cancellation between their respective X and C holes.

\section{Conclusion}\label{sec:conclusion}
In this study, we have employed the Lieb optimization approach with CCSD used as the reference wave function method to obtain accurate representations of the XC hole of the Hookium atom - a model system for which exact solutions can be obtained. Our investigation focuses on the difficulty in representing the electron-electron cusp condition within a finite Gaussian basis set, the manifestation of this in the correlation hole and effect on the cusp-related error of increasing the basis set size.
We have found that the error resulting from the cusp effect can be effectively and sufficiently reduced by using a larger basis set, and that the cusp condition in the correlation hole becomes less significant for larger $N$ Hookium atom solutions with diffuse electron densities. For smaller $N$ Hookium solutions with electron densities that are more localized, the coupling-constant-averaged XC holes were calculated using the Lieb optimization with CCSD reference wave function and used as a reference to benchmark DFT XC hole models. We confirmed the presence of significant error cancellation between the exchange hole and correlation hole for both PBE and LDA hole models and this results in their XC holes having a greater accuracy than either the exchange or correlation holes alone.

\section{Acknowledgments}
This work was supported by National Science Foundation (NSF) under Grant No. DMR-2042618. A.M.T. and T.J.P.I. are grateful for support from the European Research Council under H2020/ERC Consolidator Grant “topDFT” (Grant No. 772259). This work was supported by the Norwegian Research Council through CoE Hylleraas Centre for Quantum Molecular Sciences Grant No. 262695.

\appendix
%\section{Electron-Electron Cusp condition derivation with $s$ wave as an instance~\cite{kimball1973short}}

\iffalse
\section{Hartree-Fock Solutions for the Hookium atom}\label{app:HF_Hooke}
% I'm not sure what this section was referring to, what the Hooke's basis set is etc...
% HF with hooke's basis (basis number N=8) implemented on QUEST. The basis functions are chosen to be the harmonic-oscillator eigenfunctions\cite{o2003wave}. 
%Hartree-Fock calculations of the Hookium atom have been undertaken using a specially optimized basis set, with basis functions chosen to be the harmonic oscillator eigenfunctions~\cite{o2003wave}
\begin{equation}
    \phi_{k}(r)=\frac{H_{2 k-1}(r / \sqrt{2})}{2^{k} \sqrt{(2 k-1) ! r / \sqrt{2}}} \frac{\exp \left(-r^{2} / 4\right)}{(2 \pi)^{3 / 4}}
\end{equation}
Comparing with N=10 work~\cite{amovilli2003exact}, the N=8 basis has lower HF energy. Subsequent work has lead to an improved set of HF orbitals with a reduced the exchange energy error to $1E-7 E_h$.\cite{ragot2008comments}
\fi

% I don't understand what this means...
%What's more, the first three terms in the perturbation expansion of the exact energy and Hartree–Fock energy of the lowest singlet and triplet states\cite{gill2005electron}. Later there is a work give a better set of HF orbitals that reduced the exchange energy error to $1E-7 E_h$.\cite{ragot2008comments}

\section{Reference Potential Used in the Lieb Optimization}\label{app:ref_pot}
In this work, the reference potential employed in the Lieb optimization is a modified form of the localized Hartree-Fock potential,~\cite{Sala2001} in which the Slater non-local exchange potential is corrected at long-range by an approximate Fukui potential~\cite{Parr1984} to avoid the numerical instabilities associated with calculating the Slater potential at low densities. In terms of spin-$\sigma$ Kohn-Sham orbitals $\psi_{i\sigma}$ the Slater exchange potential $v_\text{Sx}^{\sigma}$ and approximate Fukui potential $v_\text{f}^{\sigma}$ are given respectively by
\begin{align}
    v_\text{Sx}^{\sigma}(\mathbf{r}) &= -\frac{1}{n_{\sigma}(\mathbf{r})}\sum_{ij}^{occ} \psi_{i\sigma}^{\ast}(\mathbf{r})\psi_{j\sigma}^{}(\mathbf{r}) \int \mathrm{d}\mathbf{r}'\, \frac{\psi_{i\sigma}^{\ast}(\mathbf{r}')\psi_{j\sigma}^{}(\mathbf{r}')}{\vert\mathbf{r}-\mathbf{r}'\vert} \label{eq:slater_pot} \\
    v_\text{f}^{\sigma}(\mathbf{r}) &= - \int \mathrm{d}\mathbf{r}'\, \frac{\vert\psi_{\text{HOMO}\sigma}^{}(\mathbf{r}')\vert^{2}}{\vert\mathbf{r}-\mathbf{r}'\vert}. \label{eq:fukui_pot}
\end{align}
Due to the division by density in Eq.~\eqref{eq:slater_pot} the Slater potential becomes numerically unstable to calculate in asymptotic regions where the density is very small, however the Fukui potential can be evaluated in these regions without numerical instability. In this work, the reference exchange potential is constructed from a density-dependent admixture of Slater and Fukui potentials as
\begin{equation}\label{eq:refxpot}
    \begin{aligned}
        v_\text{ref,x}^{\sigma}(\mathbf{r}) &= \kappa_{\sigma}(\mathbf{r})v_\text{Sx}^{\sigma}(\mathbf{r}) + (1-\kappa_{\sigma}(\mathbf{r}))v_\text{f}^{\sigma}(\mathbf{r}),\\
        \kappa_{\sigma}(\mathbf{r}) &= \frac{n_{\sigma}(\mathbf{r})}{\eta + n_{\sigma}(\mathbf{r})}
    \end{aligned}
\end{equation}
where the parameter $\eta$ is selected to determine the density at which the reference potential is an equal mixture of Slater and Fukui potentials - here a value of $\eta=2\times 10^{-6}$ is used. The potential in Eq.~\eqref{eq:refxpot} is used in place of the Slater potential in the calculation of the localized Hartree-Fock potential, which applies a correction to better reproduce the exact exchange potential and the result of which is used as the reference potential in the Lieb optimization. 

%\end{widetext}
\bibliography{aipsamp}
\end{document}